\documentclass[bibyear]{aa} 

\def\m{$\mu$m }

\usepackage{graphicx}
\usepackage{txfonts}
\begin{document}

   \title{The extragalactic background light revisited and the cosmic
photon-photon opacity}

\titlerunning{The Extragalactic Background Light Revisited}

   \author{Alberto Franceschini
          \and
          Giulia Rodighiero  
          }

   \institute{University of Padova, Physics and Astronomy Department,
              Vicolo Osservatorio 3, I-35122 Padova, Italy \\
              \email{alberto.franceschini@unipd.it}
             }

   \date{Received September 15, 2016; accepted March 16, 2017}

 
  \abstract
   {In addition to its relevant astrophysical and cosmological significance, the Extragalactic Background Light (EBL) is a fundamental source of opacity for cosmic high energy photons, as well as a limitation for the propagation of high-energy particles in the Universe.}
   {We review our previously published determinations of the EBL photon density in the Universe and its evolution with cosmic time, in the light of  recent surveys of IR sources at long wavelengths. }
   {We exploit deep survey observations by the Herschel Space Observatory and the Spitzer telescope, matched to optical and near-IR photometric and spectroscopic data, to re-estimate number counts and luminosity functions longwards of a few microns, and the contribution of resolved sources to the EBL. }
   {These new data indicate slightly lower photon densities in the mid- and far-infrared and sub-millimeter compared to previous determinations. This implies slightly lower cosmic opacity for photon-photon interactions.  }
   {The new data do not modify previously published EBL modeling in the UV-optical and near-IR up to several microns, while reducing the photon density at longer wavelengths. 
This improved model of the EBL alleviates some tension that had emerged in the interpretation of the highest-energy TeV observations of local \textit{blazar}s, reducing the case for new physics beyond the standard model (like violations of the Lorenz Invariance, LIV, at the highest particle energies), or for exotic astrophysics, that had sometimes been called for to explain it.
Applications of this improved EBL model on current data are considered, as well as perspectives for future instrumentation, the Cherenkov Telescope Array (CTA) in particular.}


   \maketitle
%

\section{Introduction}

The extragalactic background radiation at various electromagnetic frequencies, from radio to gamma-rays, is a fundamental constituent of the Universe, and was demonstrated to permeate it quite uniformly (e.g., Longair 2000). Such radiations have a key role during most of the history of universal expansion and the formation of cosmological structures.
Given their ubiquity, radiations are a fundamental source of opacity for the propagation of high-energy cosmic-ray particles and photons throughout space-time (Nikishov 1962; Gould and Schreder 1966).  

The most intense and cosmologically relevant among the diffuse radiation components is the Cosmic Microwave Background. As discussed below (Sect. 2), these numerous photons make a wall to the propagation of ultra-energetic particles and photons in the PeV regime (the Greisen–Zatsepin–Kuzmin effect), that has been confirmed by many experiments (see e.g., the Pierre Auger Collaboration 2010).

Another particularly important component of this radiation is the Extragalactic Background Light (EBL) in the wavelength interval between 0.1 and 1000 $\mu$m, produced by cosmic sources during the evolutionary history of the Universe. Interactions between these and very high-energy photons from astrophysical sources, and the consequent pair production, produce strong and observable opacity effects (e.g., Stecker et al. 1992, among others). The corresponding high-energy exponential cutoffs are customarily identified in the high- and very-high energy {(HE [0.2 to 100 GeV] - and VHE [above 100 GeV]) }spectra of several \textit{blazar}s observed with imaging atmospheric Cherenkov telescopes, now operating between a few tens of GeV to tens of TeV (e.g., HEGRA, HESS, MAGIC, see e.g., Dwek and Slavin 1994, Stanev and Franceschini 1998, Aharonian et al. 2006, Albert et al. 2006). Of course, not only photons from cosmic sources interact with EBL photons, but any particles of sufficiently high energy, from cosmic rays to neutrinos, in principle. Therefore, the issue of a high-precision determination of the time-evolving EBL is of central importance.

Several efforts to model the EBL and its time evolution have been published, including those based on physical prescriptions about the formation and evolution of cosmic structures, in particular the semi-analytical models of galaxy formation. This is the so-called ``forward evolution'' approach, advocated by Gilmore et al. (2012), among others. The alternatives are the more empirical approaches essentially based on local properties of cosmic sources and educated extrapolations back in cosmic time (e.g., Stanev \& Franceschini 1998, { Pei, Fall, and Hauser 1999, Stecker et al. 2006,} Stecker, Malkan \& Scully 2006, Franceschini, Rodighiero, and Vaccari 2008, henceforth AF2008), the ``backward evolution'' modeling.

Once the background photon densities are defined, the calculation of the cosmic photon-photon opacity and its effects on very high-energy spectra of \textit{blazar}s are standard physical practice (e.g., Stecker et al. 1992). Many papers have compared the predictions for such attenuated spectra with multi-wavelength \textit{blazar} observations, and the results appear often rather consistent, for both local and high-redshift objects. 
An interesting example, illustrative  of what is currently feasible, is reported in Ackermann et al. (2012), who analyzed years of Fermi Large Area Telescope
(LAT) observations of a large sample of \textit{blazar}s over a wide range of redshifts to investigate the effects of the pair-production opacity in various redshift bins, assuming smooth spectral extrapolations towards the highest energies and the EBL evolutionary models by AF2008. 
These LAT-normalized \textit{blazar} spectra show indeed cutoff features increasing with redshift just as expected by the model, providing remarkable proof of the overall validity of our understanding of the local EBL photon density and its time evolution.

Similarly, the multi-wavelength high-energy spectra of both BL Lac and Flat-spectrum radio quasars (the Active Galactic Nuclei - AGN - populations emitting at the highest energies, {blazars} henceforth) are usually fairly well reproduced in terms of standard {blazar} emission models (like the Synchrotron Self Compton or the External Compton processes), once corrected for photon-photon absorption based on the most accurate EBL models.

There are however still some areas of concern related with observations of the highest energy {blazar} spectra. 
One is the reported relative independence (or only moderate dependence) of the observed spectral indices of {blazars} in the limited redshift interval currently probed at TeV energies, while a faster increase might be expected due to the strong dependence of the opacity on the source distance (De Angelis et al. 2009; Persic \& de Angelis 2008).
Another related consideration was concerning the detection of variable emission in the energy range 70 to 400 GeV from the flat spectrum radio quasar \object{PKS 1222+216}, where such high-energy photons would be expected to be absorbed in the broad line region (Tavecchio et al. 2012).

Another source of possible concern are some apparent upturns of the spectra of {blazar}s at the highest energies, once the observed spectra are corrected for EBL absorption. A characteristic instance would have been shown by the HEGRA observations of \object{Mkn 501}, whose EBL-corrected VHE spectrum appeared to show a statistically significant hardening above 10 TeV, as reported for example in Costamante (2012, 2013).

All the above considerations may have important implications for fundamental physics, as they have raised an important question about photon propagation in space. It has been suggested that photons can oscillate into axion-like particles (ALPs), which are a generic prediction of extensions of the Standard Model of elementary particles, in particular the super-string theories. Photons can oscillate with axions in a similar way as massive neutrinos do, which would require an external field, for example a nano-Gauss magnetic field, to compensate for the photon and ALP spin difference. Then a fraction of VHE photons could escape absorption, because ALPs do not interact with ambient photons (De Angelis et al. 2007; Tavecchio et al. 2012).
%
ALPs may even be an important constituent of the cosmos because, while not contributing to dark matter, they are a possible candidate for the quintessential dark energy. 

It has also been proposed that photons and particle interactions at very high energies might reveal deviations from predictions of the Standard Model of particle physics. A violation of the Lorentz transformation of special relativity might happen, for example, in the framework of quantum gravity theories (Jacobson et al. 2003), but would only manifest itself at energies comparable to the Planck energy. As suggested by many, this violation of the Lorentz invariance (LIV) might be testable for photons of energies above $\sim$ 10 TeV interacting with the EBL, by either increasing or decreasing the standard photon-photon opacity. The famous luminous outburst of the local blazar \object{MKN 501} in 1997 has offered the opportunity to test LIV effects in the spectrum above 10 TeV, corresponding to interactions with EBL photons of wavelengths >10 \m,  if we consider the relation between interacting photon energies at the maximum of the cross section:
\begin{equation}
\lambda_{max} \simeq 1.24 (E_\gamma [TeV])\; \mu m .
\label{energy}
\end{equation}

Based on the above considerations, several experiments are now being planned or under development, targeting the highest-energy photons, such as the ASTRI mini array (Di Pierro et al. 2013; Vercellone et al. 2013), HAWC (Abeysekara et al. 2013); HiSCORE (Tluczykont et al. 2014), LHAASO (Cui et al. 2014), all well sensitive in the photon energy interval 1-100 TeV, and above. Ultimately, the Cherenkov Telescope Array (CTA) will offer vast, factor 10 sensitivity improvement compared to current instrumentation over a very wide energy range (Acharya et al. 2013).

All this calls for a revision of the EBL modeling, with a particular consideration of the longer wavelength infrared and submillimetric regime. Our previous effort in AF2008 has revealed remarkable success in reproducing absorption spectral features for sources over a wide redshift interval and up to photon energies of a few TeV. However that EBL model was only partly optimized for wavelengths above several $\mu$m,  including only a preliminary account of the \textit{Spitzer} MIPS mid-IR data.
The present paper is dedicated to reconsidering the issue of the propagation in space-time of the highest-energy cosmic photons. In particular, we update AF2008 by including new very extensive data in the far-IR and submillimeter obtained with the \textit{Herschel} Space Observatory, in a spectral region where the EBL and its evolution with time cannot be directly measured. The updated model also makes full use of deep survey data from the \textit{Spitzer} telescope.

Section 2 summarizes the new data needed for the improved EBL model. Section 3 reports on our data-fitting scheme and multi-wavelength spectral corrections. Section 4 describes the updated EBL model and the cosmic photon density. This accounts only for the contributions by known sources, like galaxies, active galactic nuclei and quasars, and does not consider various published attempts to measure the total-light EBL in the near-IR.
Section 5 summarizes our new results on the cosmic photon-photon opacities and applications to a few known {blazar}s. Section 6 contains some discussions on our results and prospects for future observations, including some detailed simulations for observations with the Cherenkov Telescope Array (CTA) concerning in particular the effects of a truly diffuse background from primeval sources.

We use in the following a standard cosmological environment with $H_0=70$ km s$^{-1}$ Mpc$^{-1}$, $\Omega_m=0.3$, $\Omega_\Lambda=0.7$.
We indicate with the symbol $S_{24}$ the flux density in Jy at 24 $\mu$m (and similarly for other wavelengths). $L_\lambda/L_\odot$ is the $\nu L_\nu$ luminosity in solar units.


\section{The new data}
\label{data}

With reference to the analysis of AF2008, the new data made available during the recent years for the estimate of the contributions of extragalactic sources to the EBL concern in particular the source's infrared emission. Instead, the available data in the optical and near-IR portion of the galaxy spectrum have largely remained unchanged. Considering also that the cosmic opacity originating from UV-optical background photons in AF2008 have proved so far quite successful in reproducing the {blazar} spectra at about 1 TeV or less, we have chosen not to modify the model at wavelengths shorter than 5 \m  (corresponding approximately to the spectral coverage of the \textit{Spitzer} IRAC surveys).

Many of the data at longer wavelengths that we will exploit in the present analysis were already discussed in detail in Franceschini et al. (2010, AF2010). We concentrate here on the new data only, that have mostly come from two infrared observatories in space, the NASA \textit{Spitzer} telescope and the ESA \textit{Herschel} observatory.

\begin{figure}[!ht]
\centering
\includegraphics[angle=0,width=0.5\textwidth,height=0.4\textwidth]{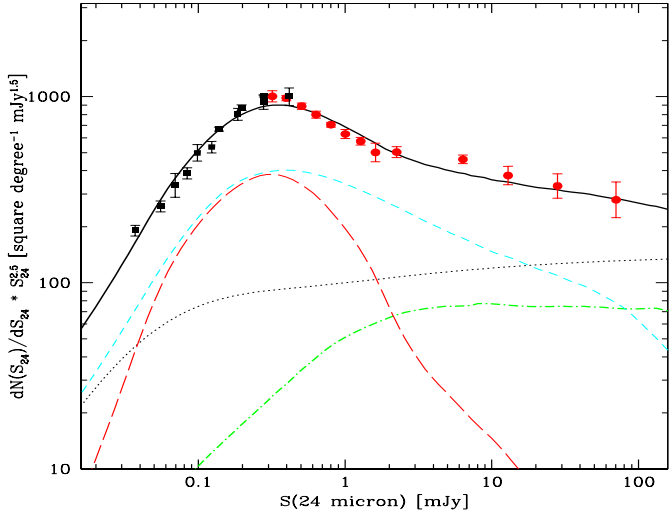}
\caption{Euclidean-normalized differential number counts of extragalactic sources at $24\mu$m compared with our model fit. 
The red circles are from the analysis of SWIRE survey data by Shupe et al. (2008), black squares from Papovich et al. (2004).
The contribution by type-I AGNs is shown as green dot-dashed line, moderate-luminosity starbursts (the LIRGs) make the cyan short-dash line (type-II AGNs and starbursts are included in the same population on the assumption that in both classes the IR spectrum is dominated by starburst emission).
The red long-dashed line corresponds to the population of high-luminosity sources dominating the IR emissivity at high redshifts.
The dotted line is the separate contribution of normal spirals, while the continuous line is the total model counts.
}
\label{c24}
\end{figure}

\begin{figure}[!ht] 
\includegraphics[angle=0,width=0.5\textwidth,height=0.4\textwidth]{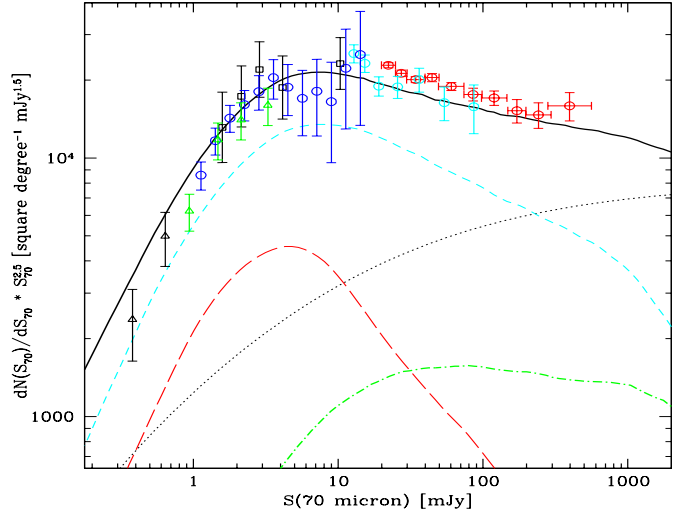}
\caption{ Euclidean-normalized differential number counts of extragalactic sources at $70\mu$m compared with our model fit.  The blue open circles are from Berta et al. (2011), black triangles from Bethermin et al. 2010). Other datapoints are as reported in AF2010.  The cyan short-dashed line is the LIRG and the red long-dashed line the ULIRG populations. Black dotted and green dot-dash lines are the normal spiral and type-1 AGN sources  (see Sect. \ref{model} for our population decomposition).
}
\label{C70EBL}
\end{figure}

\subsection{The Spitzer MIPS number counts and luminosity functions}

\begin{figure}[!ht] 
\includegraphics[angle=0,width=0.5\textwidth,height=0.4\textwidth]{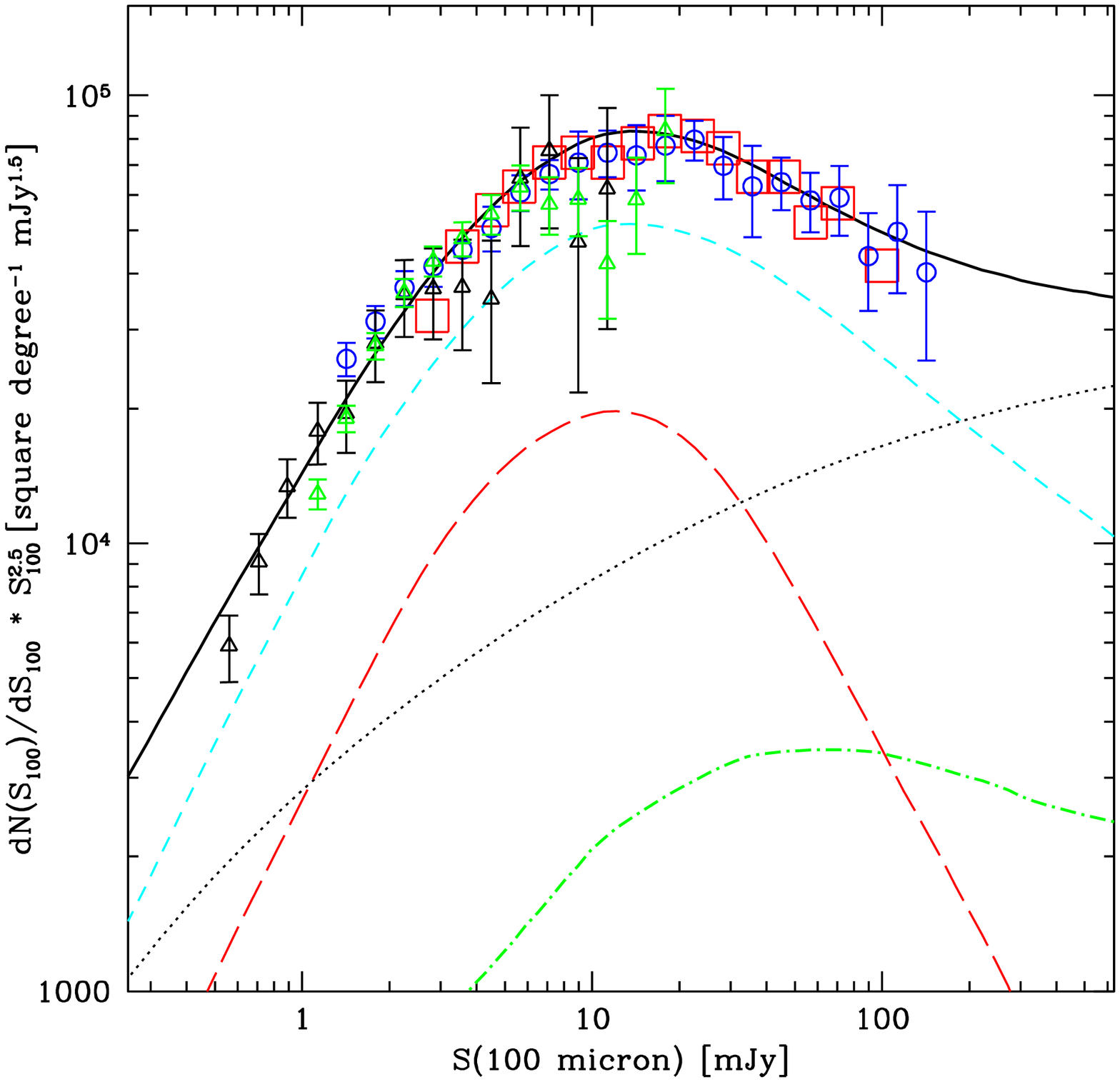}
\includegraphics[angle=0,width=0.5\textwidth,height=0.4\textwidth]{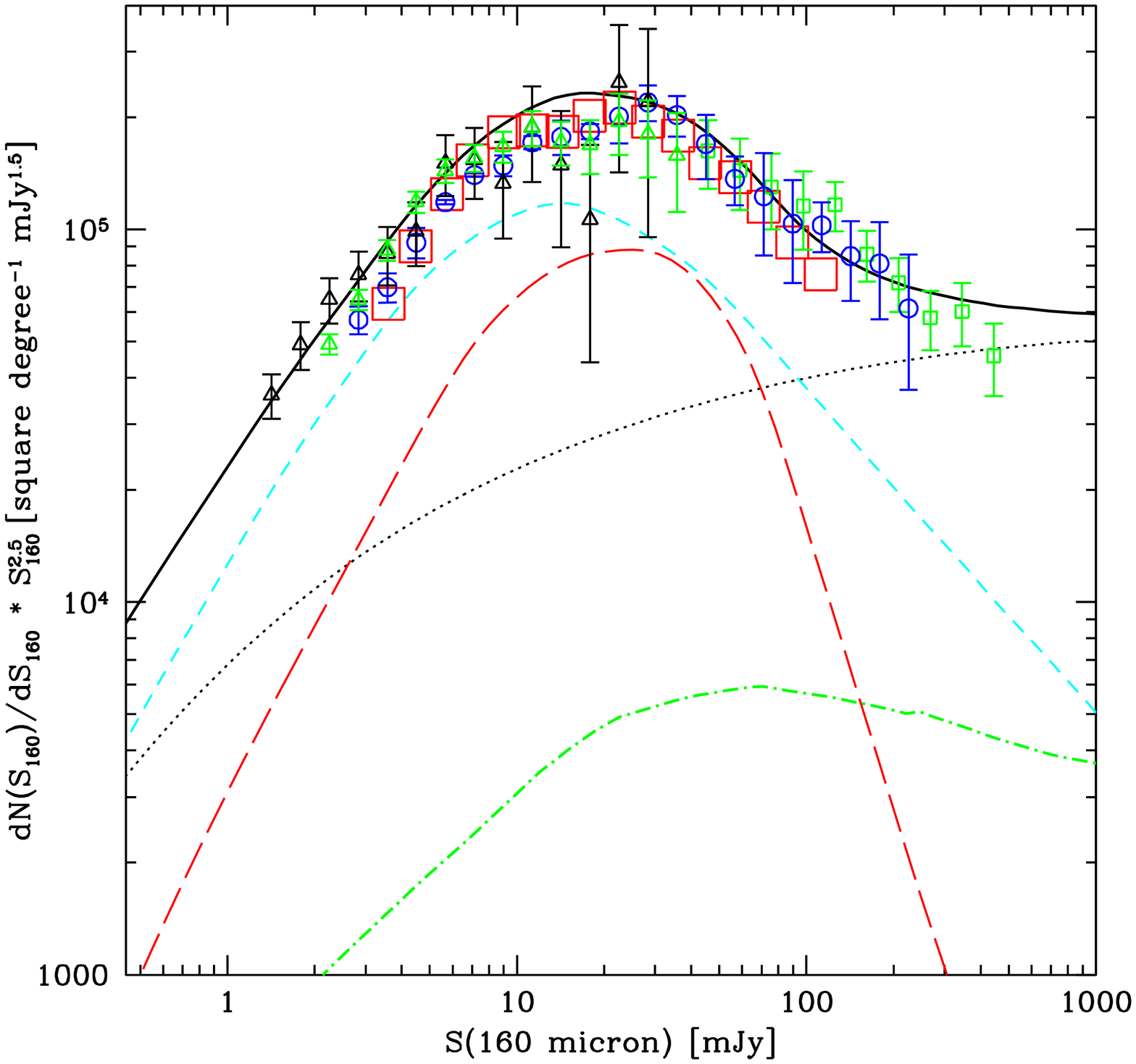}
\caption{Left: Euclidean-normalized differential number counts of extragalactic sources at $100\mu$m from Berta et al. (2010).  The cyan short-dashed line are the LIRG and the red long-dashed line the ULIRG populations. Black dotted and green dot-dash lines are the normal spiral and type-1 AGN sources.
Right: differential number counts of extragalactic sources at $160\mu$m compared with our model fit.  The red open square and blue open circle datapoints are from Berta et al. (2010) and Berta et al. (2011), respectively, while the small green open squares are a re-analysis of the Spitzer/MIPS data by Bethermin et al. (2010).  
Black triangles are from Magnelli et al. (2013).
The lines are our model fits and population decomposition (see Sect. \ref{model}).
}
\label{C100160}
\end{figure}

The \textit{Spitzer} MIPS imaging camera has been used to map several deep sky regions with the 24 $\mu$m channel and, to much shallower depths, at 70 $\mu$m.
The former observations benefited from the excellent instrumental sensitivity and good spatial resolution, while the latter observations were limited in both senses. From the local EBL viewpoint, what matters particularly are the source number counts at both wavelengths that are reported in differential normalized Euclidean units in Figures \ref{c24} and \ref{C70EBL}. 
MIPS observations at 70 \m  (and those at 160 $\mu$m) are limited by sensitivity and source confusion due to the poor angular resolution. Galaxy number counts based on MIPS data by Frayer et al. (2006) and Bethermin et al. (2010) are reported in differential Euclidean-normalized units in Figures \ref{C100160}. 
Figure \ref{C70EBL} also includes data from the \textit{Herschel} observatory that will be presented below.   All these are compared with the model predictions discussed later in the paper.

Important to note is that both these counts, as well as those discussed in following sections, show a flat, roughly Euclidean behavior at bright fluxes, a well-defined maximum at a given flux density value, and then a rather quick convergence towards faint fluxes. This has the implication that the source contribution to the total EBL intensity at the two wavelengths is mostly resolved at the limiting fluxes of the deepest surveys (see Madau \& Pozzetti 2000 for a similar consideration about the optical counts), and the residual contribution by the fainter sources is negligible. We estimate the latter, in any case, from our model fit to the data in Sect. \ref{model}, easily extrapolated to the faintest fluxes.

\begin{figure}[!ht] 
\includegraphics[angle=0,width=0.5\textwidth,height=0.6\textwidth]{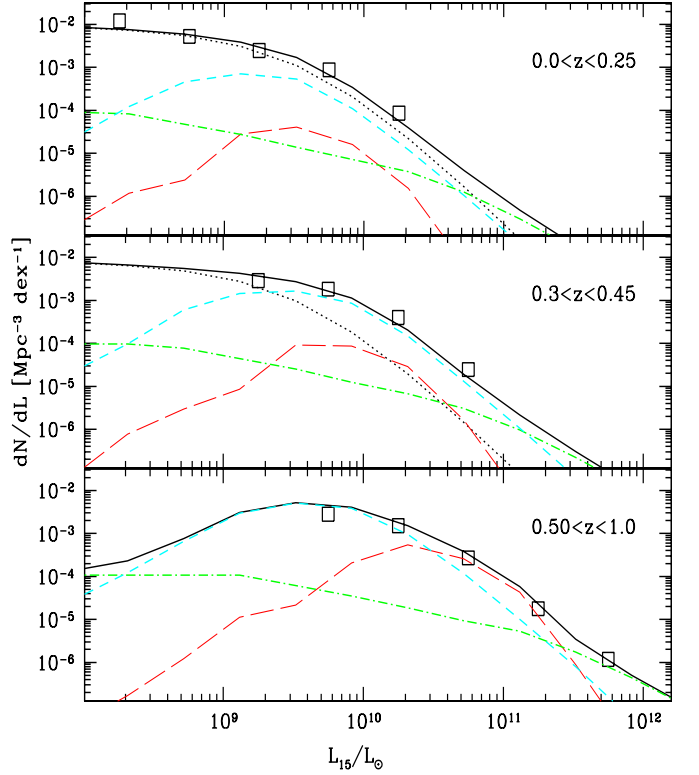}
\caption{
The rest-frame 15 \m  luminosity functions of IR-selected galaxies in redshift bins, in comoving volume units. The datapoints are from a \textit{Spitzer}-MIPS sample selected at 24 \m  by Rodighiero et al. (2010), and computed at 15 \m  to minimize the K-corrections. 
The units in abscissa are the luminosity  $\nu L(\nu)$ at 15 \m  in solar luminosities.  In ordinate these are the number $dN/dlog L$ of galaxies per cubic comoving Mpc per unit logarithmic interval of $\nu L(\nu)$.   
The lines represent our multi-population fit to the data as discussed in Section \ref{model}. In particular, green dot-dashed line: type-I AGNs; Cyan short-dashed line: evolving moderate-luminosity star-forming galaxies (LIRG); Red long-dashed line: high-luminosity starbursts (ULIRG); Lower dotted black line: quiescent spiral population. The upper black continuous line is the total predicted function.
}
\label{LF15}       
\end{figure}

Since we are interested not only in the local EBL intensity, but also in its time evolution, we need information on the source spatial density and emissivity as a function of redshift. The classically used descriptors for this are the redshift-dependent luminosity functions, giving the source number density per comoving volume as a function of luminosity and cosmic time.
Various flux-limited \textit{Spitzer} samples have been used for deriving redshift surveys with good degrees of completeness, all suitable for the calculation of redshift-dependent luminosity functions. Rodighiero et al. (2010) have worked on a complete 24 $\mu$m selected sample and used it to calculate galaxy luminosity functions at the rest-frame wavelengths of 15 $\mu$m, shorter than the observation wavelengths to minimize the K-corrections for the typical source redshift ($z\sim 1$).
We report these redshift-dependent luminosity functions in Figure \ref{LF15}, together with the predictions of our evolutionary model in Sect. \ref{model}. Functions at longer wavelengths are discussed in the following section.


\begin{figure}
  \centering
  \includegraphics[width=0.5\textwidth,height=0.4\textwidth]{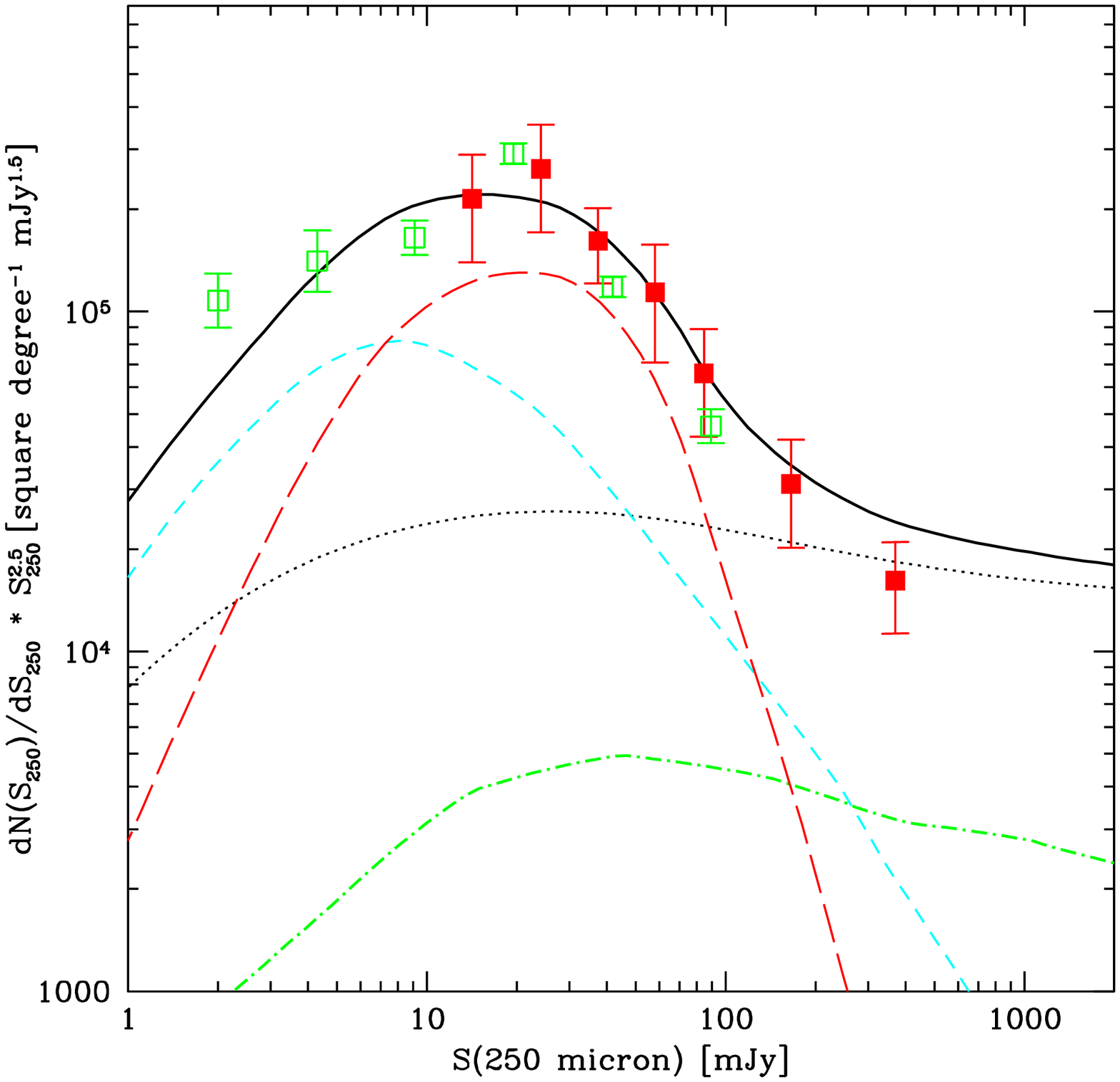}
  \includegraphics[width=0.5\textwidth,height=0.4\textwidth]{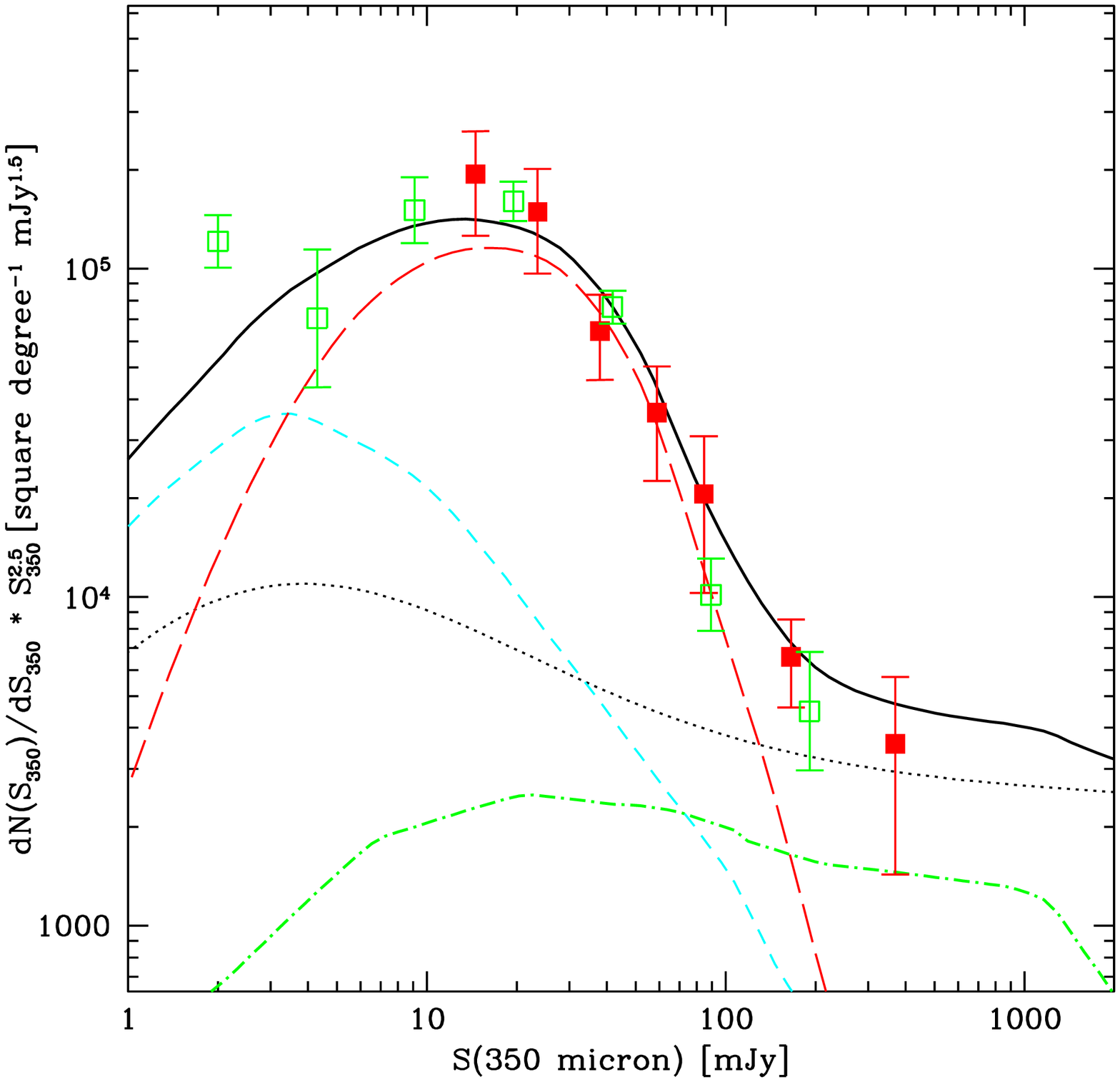}
  \includegraphics[width=0.5\textwidth,height=0.4\textwidth]{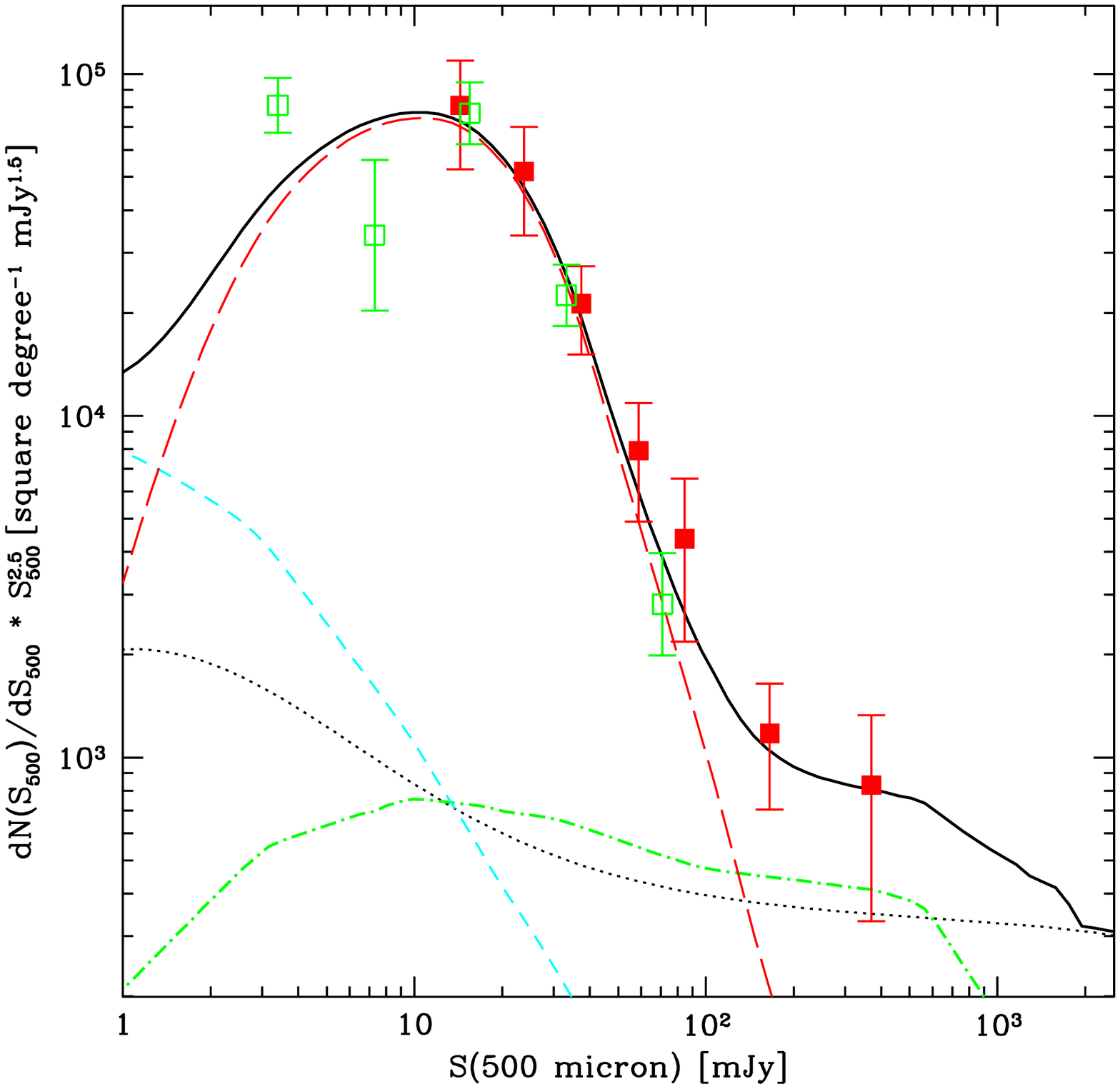}
  \caption{
From top to bottom panel: Euclidean-normalized differential number counts of extragalactic sources at 250$\mu$m, 350$\mu$m, and 500$\mu$m, from the Herschel SPIRE Hermes survey by Oliver et al. (2010). 
Green datapoints come from the P(D) fluctuation analysis by Glenn et al. (2010). 
Lines are as in the previous figures.
} 
\label{SPIRE}
\end{figure}

\subsection{The \textit{Herschel} multi-wavelength number counts and luminosity functions}
\label{herschel}

The two imaging instruments onboard \textit{Herschel}, PACS (Poglitsch et al. 2010), and SPIRE (Griffin et al. 2010), offer capabilities of deep imaging in the far-IR and sub-millimeter at the  wavelengths of 70, 100, 160 $\mu$m and 250, 350, 500 $\mu$m, respectively, with large improvement in sensitivity and angular resolution compared to previous space facilities.  
SPIRE has carried out cosmological surveys in the submillimeter over tens of square degrees below the confusion limits (the HerMES survey, Oliver et al. 2010, 2012), and over hundreds of square degrees in a contiguous fashion to sensitivities of $S_{250}\sim 30-50$ mJy (HerMES and H-ATLAS surveys, Eales et al. 2010).
%
%
PACS observations (the PEP programme, Lutz et al. 2011, and H-GOODS, Elbaz et al. 2011), in particular, have exploited the diffraction-limited imaging capability of the instrument to reduce the confusion noise and ease source de-blending and identification. PACS imaging, however, requires long integrations and has been performed to such faint fluxes only in small areas, such as GOODS and COSMOS. 

We report in Figure \ref{C100160} number count data at the effective wavelengths of 100 and 160 $\mu$m from Berta et al. (2010) and Magnelli et al. (2013), which account for all various corrections of sampling incompleteness, effective area, and fraction of spurious identifications. The 160 $\mu$m number counts include also data collected on wider areas by the \textit{\textit{Spitzer}} telescope and reported in Bethermin et al. (2010), these latter quite consistent with the \textit{Herschel}-PACS data.

The SPIRE number count data at longer wavelengths are reported in Fig. \ref{SPIRE}.
Although they are not directly related to our photon-photon opacity estimate, they are relevant to further constrain our model of the IR emissivity of cosmic sources and its evolution with redshift. 
Because we are interested in estimating absorption effects at energies up to a few tens of TeV, from eq. 1 and from consideration of the pair-production cross-section, this means that we require knowledge of the EBL from a wavelength of approximately 10 up to approximately 100 $\mu$m, that is, the interval where the EBL will never be measured because of the overwhelming brightness of the foreground interplanetary dust emission. For a precise determination of the EBL intensity, we then need to know both the local IR source emissivity and its evolution with redshift, which is constrained by the deep number counts in the far-IR and sub-mm.

As we see, in all cases, such differential normalized counts display a fast increase when going from flux densities of a few hundred mJy to $\sim$10 mJy. The slope of the counts is monotonically increasing with the effective wavelength.


The SPIRE counts are limited at the faint fluxes by the source confusion.  A significant extension to fainter limits is possible based on the analysis of the fluctuations of the background integrated intensities [the probability distribution of deflections, $P(D)$, see, for example, Franceschini 1982], as recently discussed in AF2010 and Glenn et al. (2010). 
An application to the deep \textit{Herschel} images is reported by Glenn et al. (2010), who analyzed three fields included in the HerMES programme at all three SPIRE bands (250, 350, and 500 $\mu$m). 
The differential number counts determined in this way of the  number counts are reported in Figure \ref{SPIRE} (green open squares).
These estimates are consistent with those based on individually detected SPIRE sources, and reveal clear evidence for a break in the slope of the differential counts, with a maximum and fast convergence at low flux densities.
Obviously, these data from the P(D) analysis are less reliable than those from individual source detections (in red). Nevertheless, they provide us with useful information about the convergence of the SPIRE counts at faint fluxes.

We note that only the green data points in Figure \ref{SPIRE}, and the black asterisks and red arrows in Figure \ref{EBL} below, are based on this kind of \textit{stacking} analysis; all others in the present paper are from individually detected sources at the respective wavelengths.

With this extension, the total counts in the three SPIRE bands account for as much as 64, 60, and 43\% of the far-infrared background intensity observed by COBE (red arrows in Fig. \ref{EBL}).  Indeed, in the wavelength interval covered by the SPIRE observations, the COBE/FIRAS surveys have allowed direct determination of the EBL intensity, after subtraction of the Galactic and Inter-Planetary dust emissions (Puget et al. 1996, Hauser et al. 1998, Lagache et al. 1999; see also AF2008).
All this information about the integrated number counts and the direct EBL estimates, when available, is summarized in Figure \ref{EBL}, including our updated modeling reported as a thick continuous line.
The data points and limits appearing in this figure are also reported in the review by Dwek \& Krennrich (2013), to which the reader is referred if interested in the tabulated values.

\begin{figure*}[!ht]
\includegraphics[angle=0,width=0.7\textwidth,height=0.55\textwidth]{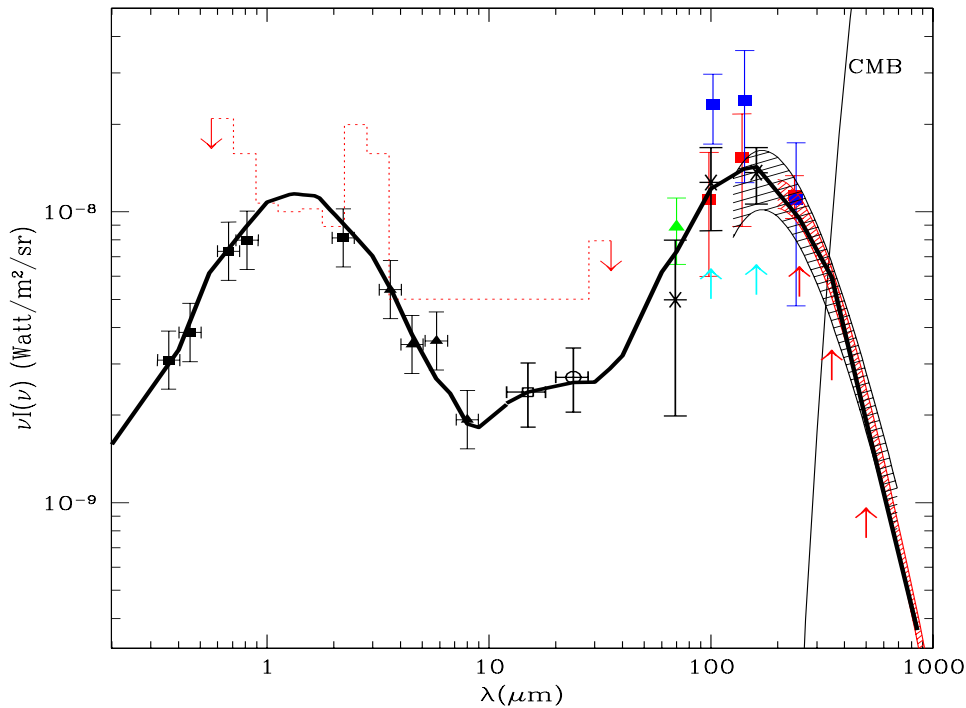}
\caption{Comparison of data on the CIRB intensity at long wavelengths and our best-fit model predictions (thick line). Most of the data are as reported and discussed in AF2008. The cyan arrowheads mark the fraction of the CIRB intensity resolved into sources by Herschel PACS (Berta et al. 2010, 2011) at 100 and 160 \m , a fraction corresponding to roughly half of the COBE intensity at these peak wavelengths. Red arrowheads at longer wavelengths mark the CIRB resolved fractions with Herschel SPIRE (Oliver et al. 2010, Glenn et al. 2010).
The three upper blue datapoints in the far-IR are from Hauser et al. (1998), the three lower (red) ones are from a re-analysis of the DIRBE data from the all-sky COBE maps by Lagache et al. (1999), the shaded areas from Fixsen et al. (1998, black shade) and Lagache et al. (2004, red shade). 
The three black asterisks at 70, 100, and 160 \m are the background intensities from PACS resolved sources and from a background fluctuation analysis by Berta et al. (2011).
The green data point comes from a background estimate from a MIPS deep survey by Frayer et al. (2006, 2009).
}
\label{EBL}
\end{figure*}

The \textit{Herschel} cosmological surveys have also allowed us to make fundamental progress towards the determination of the redshift-dependent luminosity functions at long wavelengths.
Particularly relevant are the deep multi-wavelength PACS and SPIRE observations of the GOODS-N, GOODS-S and COSMOS fields, including also deep coverage with MIPS at 70 $\mu$m. The former two cover an area of approximately 300 square arcminutes each, and the latter an area of approximately 2 sq.deg., all including a vast amount of complementary data in all e.m. bands (Giavalisco et al. 2004; Scoville et al. 2007; Sanders et al. 2007), optical spectroscopic surveys (Lilly et al. 2007) and highly precise photometric redshifts (Ilbert et al. 2010, 2013).  Berta et al. (2010, 2011) reported a detailed description of the \textit{Herschel} source extractions and data catalogs. Further deep IR surveys have been performed in the Extended Groth Strip and Extended Chandra Deep Field South for a total of more than 1000 square arcmins (Magnelli et al. 2009).

\begin{figure}[!ht] 
\includegraphics[angle=0,width=0.5\textwidth,height=0.7\textwidth]{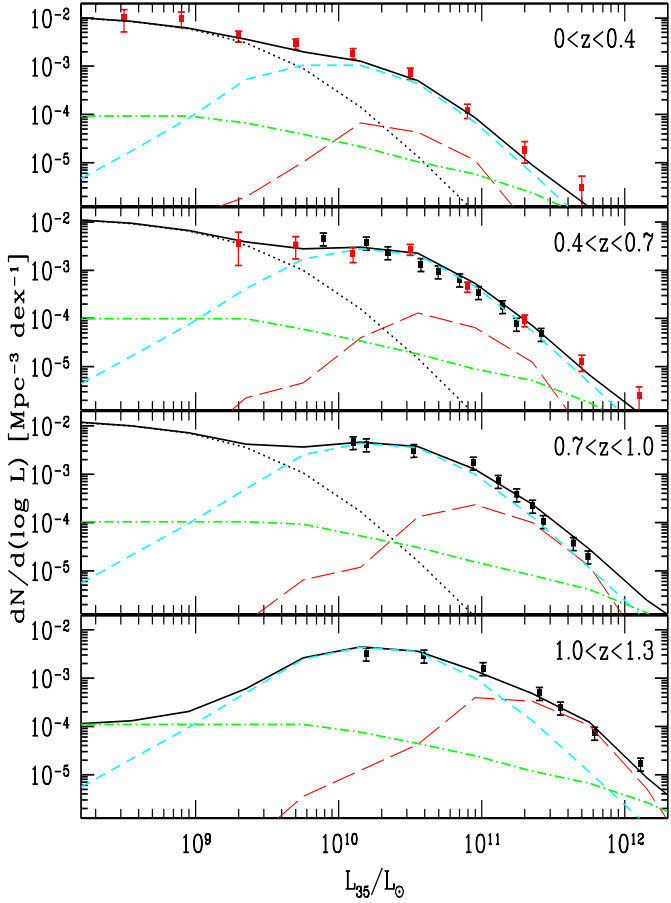}
\caption{
The rest-frame 35 \m  luminosity functions of IR-selected galaxies in redshift bins, in comoving volume units. The black datapoints are from a \textit{Spitzer}-MIPS sample selected at the effective wavelength of 70 \m  by Magnelli et al. (2009), and computed at 35 \m  to minimize the K-corrections. Red datapoints are from Gruppioni et al. (2013).
The units in ordinate and abscissa are as in Fig. \ref{LF15}.
The lines show our multi-population fit to the data (see caption to Fig. \ref{LF15}).   The upper black continuous line is the total predicted function best-fitting the luminosity function data. We limit our analysis to z=1.3 because higher-redshift sources do not contribute significantly to the EBL.
}
\label{LF35}       
\end{figure}

Gruppioni et al. (2013) and Magnelli et al. (2009) published independent multi-wavelength luminosity functions at the rest-frame 35 $\mu$m wavelength, over a wide redshift interval from local to $z\simeq 4$, well consistent with one another. 
These results are reported in Figure \ref{LF35} as red datapoints from Gruppioni et al. (2013) and black datapoints from Magnelli et al. (2013), both compared  with our model fit.

Luminosity functions at longer wavelengths were also calculated by many authors based on the deep fields (Gruppioni et al. 2010, 2013; Magnelli et al. 2013; Eales et al. 2010).
In addition, \textit{Herschel} SPIRE data over a total of 39 deg$^2$ within five high-latitude fields have been used by Marchetti et al. (2016) to compute luminosity functions at 250/350/500 $\mu$m, and provide us with a census of the luminosity density in the low-redshift Universe at these wavelengths.

All these data are relevant to constrain our models of the IR source emissivity discussed in the following Section.

\section{The data-fitting scheme}
\label{model}

To interpret the previously discussed statistical data, we have adopted the multi-wavelength modeling scheme of IR galaxy evolution reported in AF2010 and used in AF2008, with few changes and some parameter optimization. 
Our aim consists of obtaining a phenomenological fitting scheme with the best possible adherence to this large variety of data in order to use it to accurately quantify the source IR emissivity. For this reason, and considering its good ability to reproduce the data, we did not attempt to substantially update our evolutionary model, by, for example removing our distinction between LIRG and ULIRG populations (see below) that might be seen as somewhat schematic.

\subsection{Population components}
\label{pop}

Our model considers four population components characterized by different physical and evolutionary properties. 
The first class are \textbf{normal spirals}, dominating the local source populations and assumed to contribute only at $z<1$. 
The presence of this non-evolving population has already  been identified in AF2010 in order to explain the observed number counts at bright fluxes and the low-redshift luminosity functions.

A second class, already introduced by Franceschini et al. (2001) among others, is a population of \textbf{star-forming galaxies of moderate luminosities} (LIRG). In order to explain the fainter number counts at all IR wavelengths, this population needs to evolve fast in redshift up to $z\sim 1$. The LIRG's evolution rates in both luminosity and number density are slightly modified compared to AF2010 to account for the new Herschel data.
The local fraction of the moderately-luminous star-forming population is assumed to be $\sim$ 10\% of the total galaxy population, in line with the observational constraints.

A third evolutionary population introduced by AF2010 to explain the Spitzer 24 $\mu$m and SCUBA data at high-redshifts are the \textbf{ultra-luminous infrared galaxies (ULIRGs)}, dominating the cosmic IR emissivity above $z\simeq 1.5$ and, as such, less relevant for our present analysis. 
This class of sources is essential to reproduce the statistics, particularly the number counts, in the submillimeter (Herschel) and millimeter (JCMT, IRAM, CSO, APEX).
We note that, as explained in AF2010, our used terminology for LIRG and ULIRG 
indicates objects of approximately $10^{11}\ L_\odot$ at $z\sim 1$ and $\sim 10^{12}\ L_\odot$ at $z > 1.5, $ respectively.

Finally, we consider the contributions of \textbf{AGNs}, in particular type-I AGNs that are easily identified with simple combinations of optical-IR colors (Polletta et al. 2006, Hatziminaoglou et al. 2005).  The type-II category, instead, is much more difficult to disentangle among the star forming galaxy population, and we do not treat them separately here.

Our choice to distinguish two separate populations of star forming galaxies, LIRGs and ULIRGs, with different luminosity functions and evolutionary histories, was suggested  in order to fit the large observational dataset in the IR, that appear to require at least two components, one of moderate luminosity dominating at low $z$ and a second high-luminosity class at higher redshifts. This choice allowed us the flexibility to fit statistical data in the mid-IR (e.g., the 24 $\mu$m counts) and in the sub-millimeter simultaneously. Simpler realizations, for example, including only one component of star forming galaxies, would not allow us to achieve acceptable fits (see also M. Rowan-Robinson 2009).

\begin{figure}[!ht] 
\includegraphics[angle=0,width=0.5\textwidth,height=0.45\textwidth]{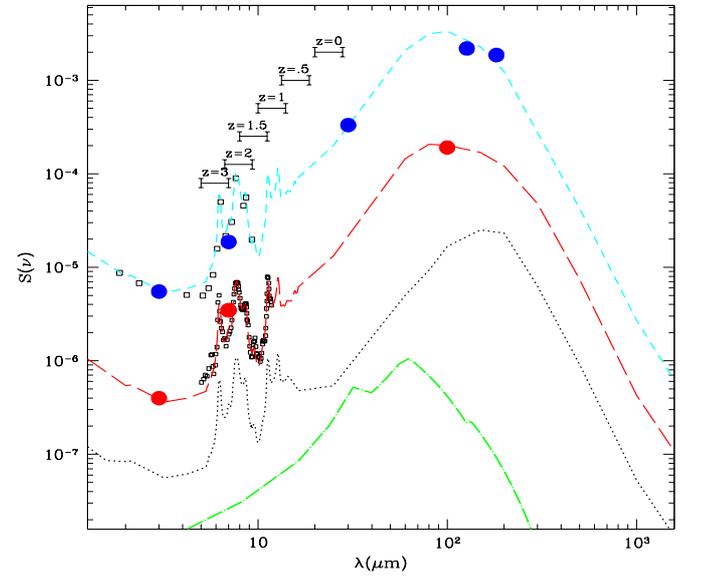}
\caption{Our adopted IR spectra of various galaxy populations.  The short-dashed cyan line corresponds to our adopted spectrum for the moderate-luminosity LIRG star-forming population, while the red long-dashed curve is the spectrum of high-luminosity ULIRG sources. In both cases the spectra are similar to that of the prototype star forming galaxy M82 (in the range from 5 to 18 $\mu m$ it is precisely the ISOCAM CVF spectrum of \object{M82}). 
The lower dotted line corresponds to a low-luminosity inactive spiral (see Sect. \ref{spectral} and AF2010), while the upper dotted line is closer to that of ULIRGs.
The lower dot-dash green line is the average type-I AGN spectrum. The boundaries of the MIPS 24 $\mu$m filter are also shown in the source rest-frames at various source redshifts.
The open squares are the average fluxes estimated by Fadda et al. (2010) for a faint sample of LIRGs at $z\sim1$ and ULIRGs at $z\sim2$.
The two blue filled circles at 127 and 182 $\mu$m and the red filled datapoint at 100 $\mu$m are the average fluxes for LIRGs and ULIRGs from \textit{Herschel} observations.}
\label{spectrum}
\end{figure}

\subsection{The spectral model}
\label{spectral}

In consideration of the multi-wavelength nature of the data used, a key ingredient in our description of the IR-selected galaxy emission is the spectral model to be adopted for the various considered sources. This is needed to calculate both the K-corrections and the transformations and interpolations of the luminosity functions at various wavelengths, as well as their volume emissivities. 

Since AGNs of type-I mostly contribute in the mid-IR, and have little influence on the far-IR and sub-mm statistics, we have simply imposed \textit{a priori} the spectral shape for them. We assume for type-I AGNs a spectral energy distribution (SED)  corresponding to an emission model by a pure face-on dusty torus from Fritz et al. (2006), as reported with a green line in Figure \ref{spectrum}. 

As for the normal spiral population, we have simplified the treatment by AF2010 of a luminosity dependent spectral shape, by assuming a single constant spectrum, shown as a dotted line in Figure \ref{spectrum}, and dominated by cold dust.

Concerning the LIRGs and ULIRG populations, modeling their IR spectra is critical to achieve good fits to the data. Our adopted SEDs for both classes, that we assumed to be independent of redshift and luminosity, are reported in Figure \ref{spectrum} as the cyan and red dashed lines. They have been obtained by slightly modifying the spectrum of the starburst galaxy \object{M82} (see e.g., Polletta et al. 2007): while in the range from 5 to 18 $\mu$m they are fixed to the mid-IR spectrum of M82 observed by ISOCAM (Franceschini et al. 2001), at longer wavelengths the two spectra have been modified so as to best-fit all the multi-wavelength data reported in Section \ref{data}.
The two average SEDs for LIRGs and ULIRGs needed for reproducing the data are not very dissimilar to one another, but the ULIRG spectrum appears to be broader, with a significant enhancement at $\lambda\sim 60 \ \mu$m and in the submillimeter. This is likely due to significantly different physical conditions in galaxies at different epochs.



Now, it is important to confirm our adopted spectral shapes with further direct observational constraints.  Fadda et al. (2010) published ultra-deep mid-IR spectra of a representative and complete sample of 48 infrared-luminous galaxies obtained with the \textit{Spitzer} IRS spectrograph. Half of these are LIRGs at $z<1$ and the other half are ULIRGs at $z>1$. 
The average spectra for the two classes of sources, as obtained by Fadda et al., are reported in Figure \ref{spectrum} as small open squares, covering the ranges of 4-10 and 5-12 $\mu$m.  As already remarked by Fadda et al. and shown in the figure, \textit{Spitzer} IRS observations are in good agreement with our adopted mid-IR spectral shapes for both LIRGs and ULIRGs.

To complement the \textit{Spitzer}'s data with longer wavelength data, Lo Faro et al. (2013) analyzed deep \textit{Herschel} SPIRE and PACS photometric data in the GOODS-South field for 10 LIRGs and 21 ULIRGs of the Fadda et al. sample. 
We show in Figure \ref{spectrum} the measurement of the average flux for the two subsamples as the blue filled circles for the LIRGs and red filled circles at 100 $\mu$m for the ULIRGs.
Overall, we find excellent agreement between these average fluxes and our synthetic spectra, which supports our adopted spectral model.



Our model luminosity functions are calibrated to primarily fit those estimated with \textit{Herschel} in the far-IR and sub-mm, as discussed in Section \ref{data}. The spectral model is particularly critical to reproduce well LFs in the mid-IR, like those from the IRAS 12 $\mu$m survey, and in the mid- and far-IR by \textit{Spitzer} and \textit{Herschel}, that are reported above.

\begin{figure}[!ht] 
\includegraphics[angle=0,width=0.5\textwidth,height=0.45\textwidth]{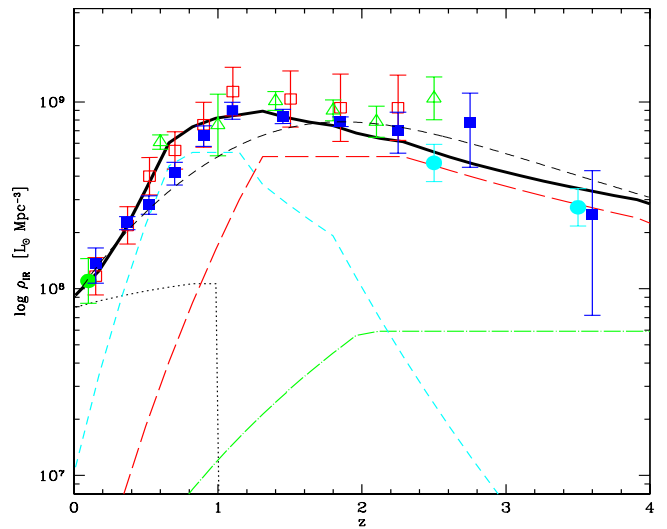}
\caption{
Data on the comoving IR bolometric luminosity density from 8 to 1000 $\mu$m as a function of redshift for the IR-selected galaxy population, compared with our model prediction discussed in Sect. \ref{model}. The luminosity density here is expressed in solar luminosities per cubic Mpc.  
Line types are as in the previous figures. 
Blue filled squares are estimates of the bolometric IR emissivity based on the \textit{Herschel} data by Gruppioni et al. (2013) and the cyan circles by Dunlop et al. (2017).
Red filled square datapoints are from the analysis of the 24 $\mu$m luminosity functions by Rodighiero et al. (2010), green triangles are from Perez-Gonzalez et al. (2005), all expressed in terms of bolometric luminosity density (left ordinate axis). The filled green circles are from Vaccari et al. (2010) and Marchetti et al. (2016). 
The black dashed line corresponds to the Madau \& Dickinson (2014) fit to the star-formation rate density versus redshift, scaled from $M_\odot/yr$ to IR luminosity in $L_\odot$ with the factor $1.7\times 10^{-10}$, as in Kennicutt (1998) for their adopted Salpeter's IMF.
}
\label{sfrEBL}
\end{figure}

\subsection{Model to data comparison}

A partial comparison of our model fitting with the statistical dataset is reported as thick lines in Figure \ref{c24} to \ref{LF35}.  
As we see, the fits are from `acceptable' to `excellent' in all cases. The same happens to data at longer wavelengths, that are less relevant to our analysis however.

For completeness, we also summarize in Figure \ref{sfrEBL} our best guess evolution model for the time-dependence of the comoving IR galaxy emissivity (black continuous line), compared with existing literature data, where we see a fast increase of it from z=0 to 1. 
Our estimated IR luminosity density shown as a thick black line uses all IR data reported in the review by Madau \& Dickinson (2014), plus some other independent determinations. The black dashed line in the Figure is their best-fit to the star-formation rate density translated back to IR bolometric luminosity with the Kennicutt (1998) recipes (see Figure caption).
Further to a fair overall match between our analysis and Madau \& Dickinson's, there are some small differences that are apparent in the figure, in particular our estimate is slightly higher than theirs at $0.5<z<1.5$ (and slightly lower above). This results from the fact that Madau \& Dickinson's fit, trying to  intercept both the IR and UV data points, suffers some tension between the two sets of data (the IR density shows a maximum at slightly lower redshifts than the UV, even corrected for dust extinction). Our analysis reproduces more closely the IR data of our interest here.

The result reported in Figure \ref{sfrEBL} is important to further constrain our modeled EBL intensity between 10 and 300 \m .

While we have adopted a standard $\chi^2$ test to evaluate the goodness-of-fit to individual datasets (like e.g., the number counts in a given waveband or the luminosity function at a rest-frame wavelength and redshift) and to verify the occurrence of possible serious misfits, 
we have not attempted to run global $\chi^2$ minimization and automatic $\chi^2$ searches of the model parameter space to look for degeneracies in the solutions and to provide quantitatively defined uncertainty intervals. This was because of the huge number of datapoints with largely inhomogeneous error estimates, and hidden and unaccounted-for systematic uncertainties in specific data. The latter is particularly the case for the redshift-dependent LFs, for which a reliable error estimate free of systematic uncertainties is virtually impossible. In some other cases, like for the FIRAS-CIRB intensity (see Fig. \ref{EBL}), errors are not even defined except for an uncertainty range, not suited for a $\chi^2$ test. The consequence is that we will not offer in the following a real discussion of uncertainties and parameter degeneracies.
Note that our analysis benefits from an extreme adherence to a vast amount of data, at the cost of an incapacity to provide a well-defined quantitative confidence range. 
However, considering only the constraints set by the source multi-wavelength number counts, we can estimate a global uncertainty of about $\pm$10\% on them, which directly translates onto a $\pm$10\% uncertainty on the photon optical depth up to z$\simeq$1 and increasing above. The linear relation between number counts, background intensity, and photon-photon optical depth can be seen in eqs. 5 to 13 in AF2008.


\section{EBL model improvements and the cosmic photon density}

Our current modification of the EBL model by AF2008 is based on new data at long wavelengths, $\lambda > 10$ \m  obtained from the \textit{Herschel} deep and wide-area surveys, improving previous determinations based on the \textit{Spitzer} data. Indeed, the latter required some substantial extrapolations from the observed mid-IR to longer wavelengths, that are not confirmed by the new far-IR data. Elbaz et al. (2010) and Rodighiero et al. (2010) found, in particular, that the highest-luminosity galaxies over a wide redshift interval have a relatively lower IR luminosity with respect to previous estimates. Consequently, both the new estimated contribution of sources to the EBL intensity at IR wavelengths and the photon number densities at $\lambda > 8\mu$m are slightly lower than reported by AF2008.

Figure \ref{EBL} shows our current assessment of the local EBL intensity (background intensity at redshift $z=0$) after the above mentioned modifications. We see that all the spectral range from the UV to approximately 8 \m  has remained unchanged as a consequence of the unchanged modeling of the galaxy populations at these wavelengths. Instead, the mid-IR portion from $\sim$8 to 40 \m  is now lower because of the slightly reduced  IR emissivity of galaxies. More significant is the decrease of the whole far-IR and submillimetric peak of dust emission from galaxies; the peak EBL emission at 160 $\mu$m, while estimated to exceed 20 $nW/m^2/sr$ by AF2008 based on \textit{Spitzer} data, is now lowered to $\nu I(\nu)\sim\ 14\ nW/m^2/sr$, in good agreement with several other independent determinations (e.g., Lagache et al. 1999; Berta et al. 2011).

\begin{figure}[!ht] 
\includegraphics[angle=0,width=0.5\textwidth,height=0.45\textwidth]{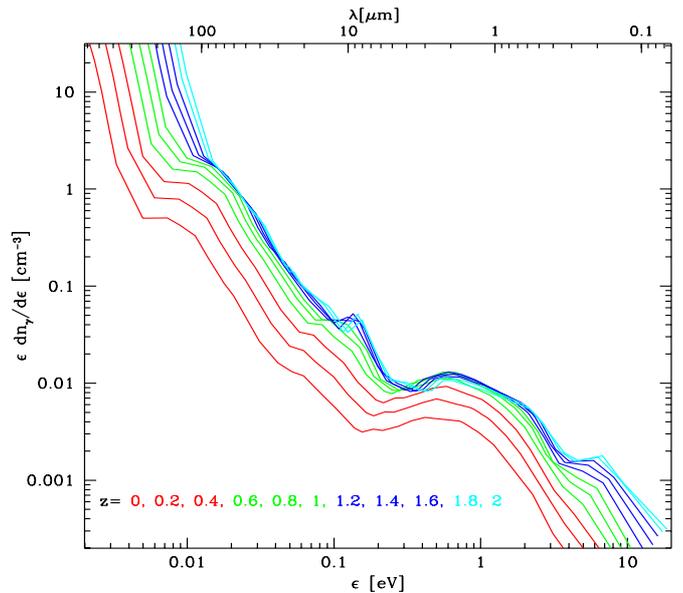}
\caption{
The proper number density of EBL photons (energy-weighted) as a function of the energy $\epsilon$. 
The various curves correspond to different redshifts, as indicated in the figure insert.
 }
\label{EBLnd}
\end{figure}

As already stressed, our work accounts only for the known source contributions, like those from galaxies, active galactic nuclei and quasars. 
We do not attempt to consider here various published measurements of the total-light EBL in the near-IR based on COBE (Gorjian et al. 2000, Wright \& Reese 2000, Cambresy et al. 2001, Hauser \& Dwek 2001, Dwek \& Krennrich 2005, 2013), and IRTS (Matsumoto et al. 2005; Kashlinsky 2006) observations, and rocket experiments (Zemkov et al. 2014).
These measurements suffer substantial uncertainties due to the poorly understood contribution of the Zodiacal Sun-reflected light.
The occurrence of a truly diffuse additional component will only be discussed in Sect. 6 as signals potentially reachable by future observations with Cherenkov telescopes.

Unfortunately, the whole interval from 24 to 70 \m  was not covered by any one of the space observatory missions during the last 20 years (ISO, Spitzer, Herschel), and this is reflected in the lack of the corresponding data in Fig. \ref{EBL}. However, thanks to our spectral analysis in Sect. \ref{spectral}, the excellent available information at 24, 70, and 100 \m  is easily interpolated inside that wavelength range, as we have done for the luminosity functions in Figure \ref{LF35}. Note that our local EBL model at $\lambda=70\ \mu$m fits well the two published independent estimates in Figure \ref{EBL}.
In conclusion, we are completely confident that we control in an accurate and robust way the background intensity produced by sources over the whole IR domain, both the local EBL and its time evolution.

Figure \ref{EBLnd} illustrates our model prediction for the time-dependent number density of diffuse cosmic photons. Again the difference with AF2008 is confined to photon energies $\epsilon < 0.2\ eV$. At the energy of 0.01 $eV$, corresponding to the peak of the photon density produced by cosmic sources (see Fig. 4 in AF2008), this difference is approximately a factor 2.5, in the sense of lower density values from the present analysis.
For completeness, in our Figure \ref{EBLnd} we have added the contribution of the Cosmic Microwave Background (CMB) photons, showing up at wavelengths $\lambda >100\ \mu$m and $\epsilon < 0.01\ eV$ as a very steep rise in photon density with wavelength and redshift.

\begin{figure}[!ht] 
\includegraphics[angle=0,width=0.5\textwidth,height=0.55\textwidth]{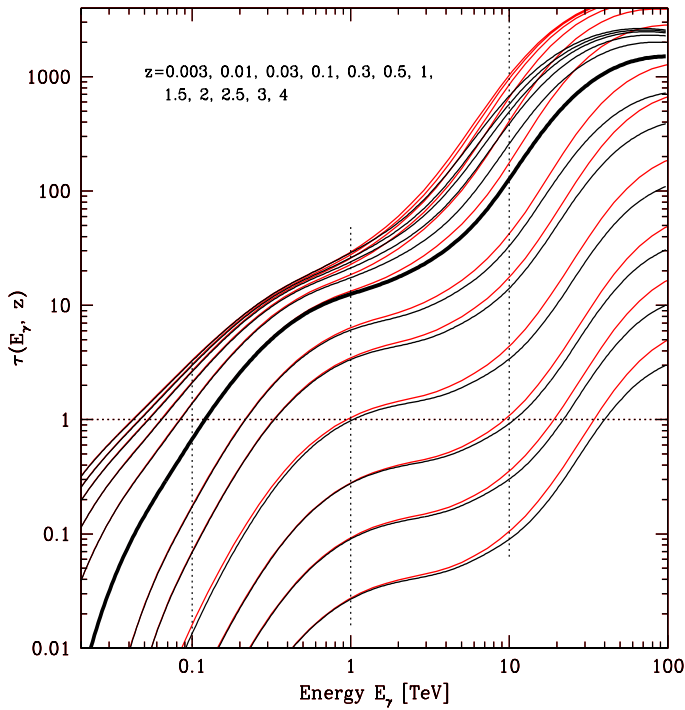}
\caption{
The optical depth by photon-photon collision as a function of the photon energy for sources located at z = 0.003, 0.01,
0.03, 0.1, 0.3, 0.5, 1, 1.5, 2, 2.5, 3, 4, from bottom to top. Black and red lines are from the present work and from AF2008, respectively. 
For the reader's convenience, the thick black line marks the z=1 line.}
\label{tauEEBL}
\end{figure}

\begin{figure}[!ht] 
\includegraphics[angle=0,width=0.5\textwidth,height=0.55\textwidth]{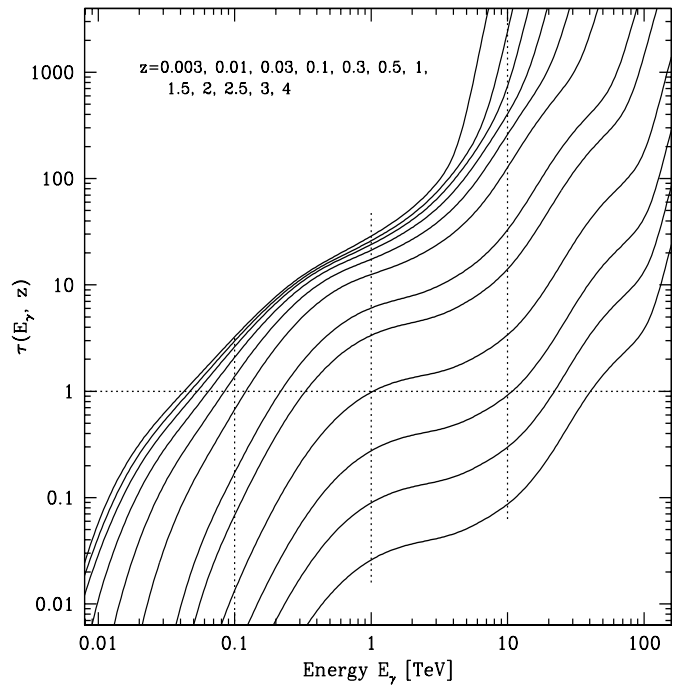}
\caption{
Same as in Fig. \ref{tauEEBL}, but including the contributions of CMB photons, producing a fast increase at the highest photon energies.
Redshifts increase from bottom to top figure.
}
\label{tauEEnew}
\end{figure}

\begin{figure*}[!ht] 
\includegraphics[angle=0,width=0.9\textwidth,height=0.9\textwidth]{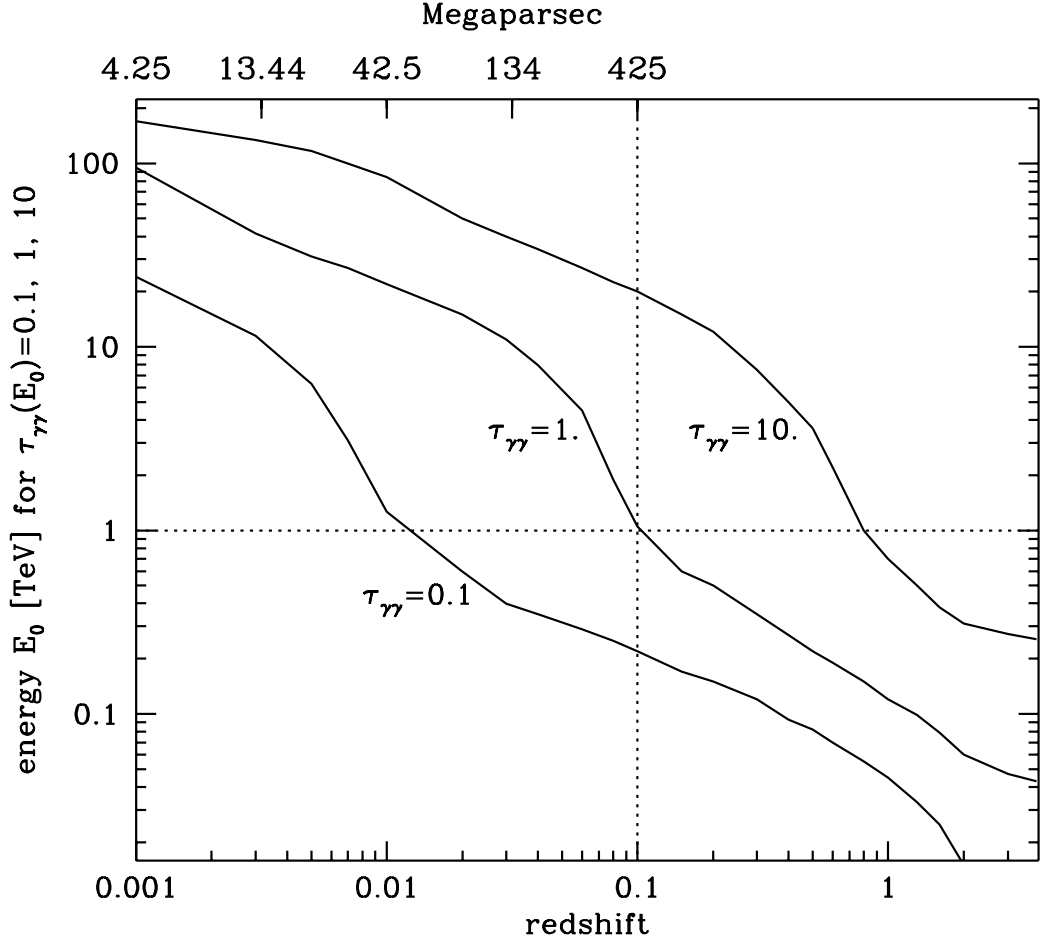}
\caption{
The energies corresponding to optical depth values of $\tau= 0.1, 1$ and 10 for photon-photon collisions, as a function of the redshift distance of the source. 
}
\label{tau1}
\end{figure*}

\section{The cosmic photon-photon opacity}

Our previously discussed modifications of the local EBL intensity and its evolution impact on the estimate of the cosmic opacity for photon-photon interaction and pair-production, and on the HE and VHE photon horizon. 
All detailed formalism and calculations of the optical depths as a function of photon energies and cosmic distances are reported in AF2008, to which we defer the reader.

The detailed dependences of the photon-photon optical depth $\tau_{\gamma\gamma}$ on the photon energy for sources over a range of distances are reported in Figure \ref{tauEEBL}, and compared there to those calculated by AF2008. 
As we see, differences due to our modified EBL concern mostly very high-energy observations of cosmic sources, from approximately 1 TeV to few tens of TeV photon energies, in agreement with the usual rule-of-thumb of  Eq. \ref{energy}, where our currently reduced background photon densities at long wavelengths imply reduced opacities. 
At lower energies the differences with AF2008 become essentially null.

The effects of the inclusion of CMB photons in the optical depth calculations are instead illustrated in Figure \ref{tauEEnew}. These effects start to be prominent at $E_\gamma\geq 100$ TeV for local sources, and at progressively lower energies at increasing source distances.

The cosmological horizons for different reference values of the photon-photon optical depth $\tau_{\gamma\gamma}$ and of the photon energy are reported in Figure \ref{tau1}.
At an energy of $E_0\simeq 1\ TeV$ there is a maximum rate of the decrease of $E_0$ with redshift for a fixed $\tau$ because of the maximum in the EBL at $\lambda \sim 1\ \mu$m. The flattening at the highest $E_0$ values is due to the CMB, while that at the highest redshifts is due to the decreased density of photons produced by galaxies at $z>1$.

Based on the constraints from the number counts only, we estimate at roughly $\pm$10\% the uncertainties on $\tau(E_\gamma,z)$ for $z<1$, and larger for $z\geq 1$.

\begin{figure*}
\centering
\begin{minipage}{0.35\textheight}
\rotatebox{0}{\resizebox{9.cm}{!}{
\includegraphics{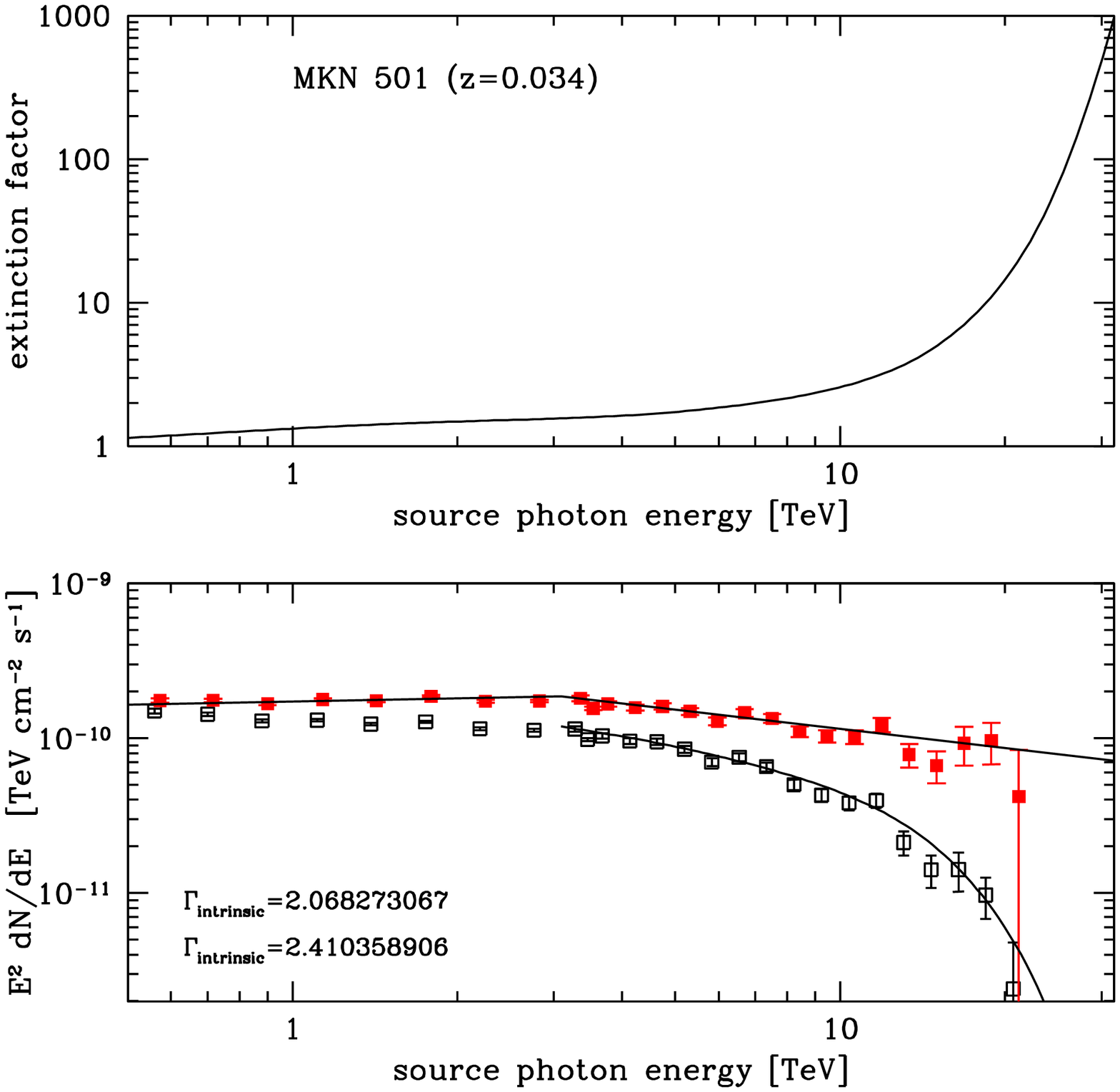}}} 
\end{minipage}
\begin{minipage}{0.35\textheight}
\resizebox{9.cm}{!}{
\includegraphics{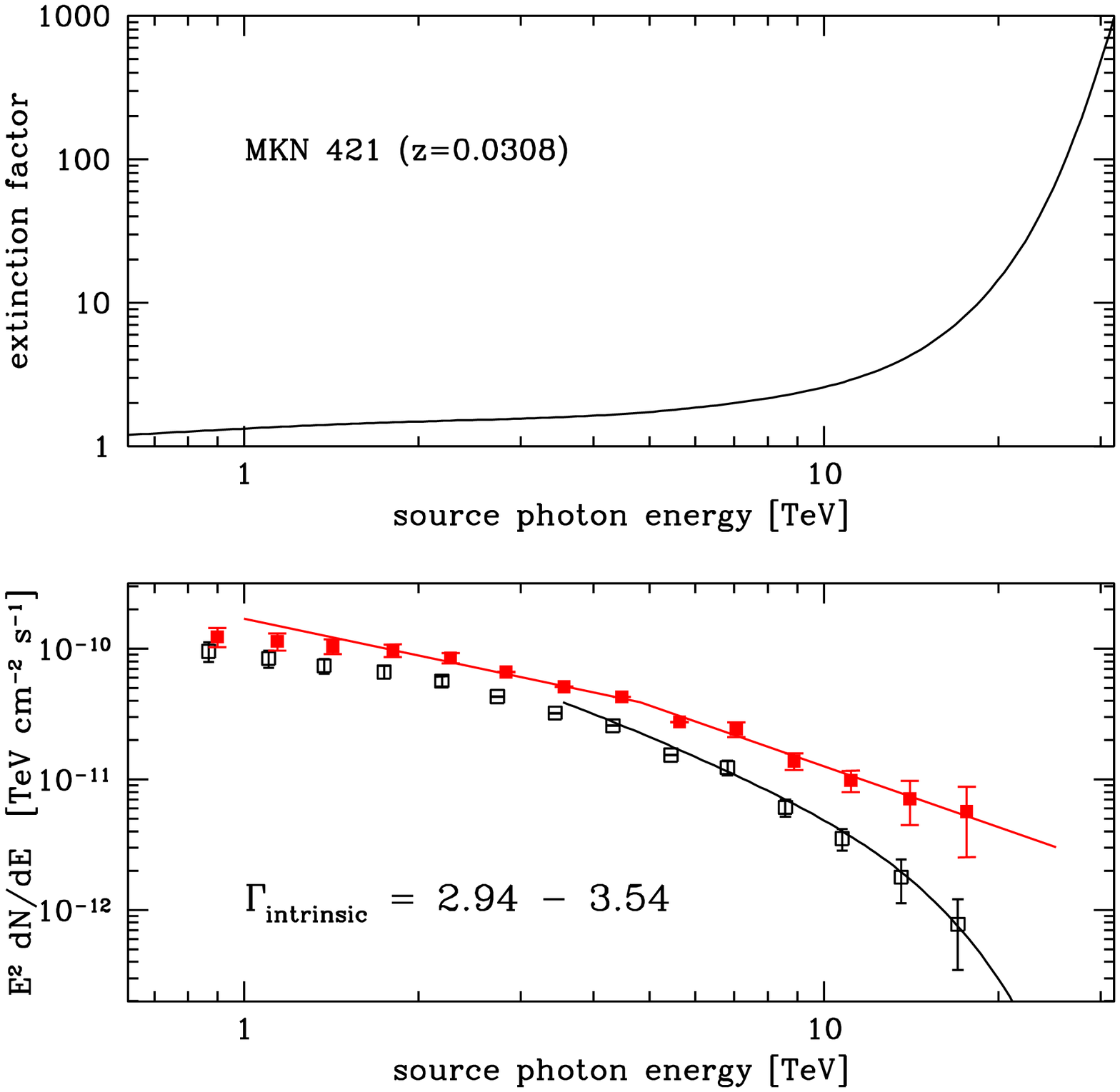}}
\end{minipage}
\caption{\textit{Top left: } The photon-photon absorption correction ($exp[\tau_{\gamma\gamma}]$) 
for the source \object{MKN 501} at $z=0.034$, based on our improved EBL model.
\textit{Bottom left:} The observed (open black) and absorption-corrected (filled red) spectrum. Data are taken from Aharonian et al. (2001, see also Aharonian et al. 1999). The intrinsic spectrum is fitted by a broken power-law with relatively flat photon spectral indices, as indicated in the insert.
\textit{Top right: } Absorption correction for the source \object{MKN 421} at $z=0.03$ .
\textit{Bottom right:} The observed (open black) and absorption-corrected (filled red) spectrum. Data are taken from Aharonian et al. (2002). The intrinsic spectrum was fitted by a steep broken power-law. Spectral indices are indicated.} 
\label{mkn}
\end{figure*}

Our improved EBL modeling at the long IR wavelengths is particularly relevant for the interpretation of the highest-energy observations of {blazar}s with Cherenkov telescopes.
Because of the strong dependence of $\tau_{\gamma\gamma}$ on distance, such high-energy multi-TeV photons can be better observed from local or low-redshift sources. Data on the two best known such examples, \object{Markarian 501} (Mkn 501, z=0.034) and \object{Markarian 421} (Mkn 421, z=0.031), are reported in Figure \ref{mkn}. In the top panels,
our best-guess extinction factors $e^\tau_{\gamma\gamma}$ 
corresponding to the two source distances are reported as a function of the photon energy.
In the bottom panels, we compare the observed (black) and extinction-corrected (red) spectral data, together with the corrected best-fit slopes.
The data on the former source, particularly relevant, come from a re-analysis of HEGRA atmospheric Cherenkov observations of the famous 1997 flares of \object{Mkn 501}, based on algorithms providing improved energy resolution to best constrain the highest energy portion of the spectrum (Aharonian et al. 2001). 
As for the Mkn 421 data, these were taken during the strong TeV gamma-ray outbursts registered during the observational season of 2000-2001 (Aharonian et al. 2002). 

In neither case do we see evidence in the absorption-corrected spectra for excess flux or upturns at the highest energies that might be indicative of an improper spectral correction. 
Some high-energy upturns for the two sources that were indicated by the analysis in AF2008 (their Fig. 9), and more significantly in Costamante (2013), are not confirmed by the present analysis.
While the differences in $\tau_{\gamma\gamma}$ between this latter and those reported in AF2008 are small at such low redshifts (see Fig. \ref{tauEEBL}), because of the exponential dependence of the absorption correction on $\tau$, these moderately significant upturns are removed.

Aharonian et al. (2002) have performed a systematic comparison of the VHE spectra of the two sources and, based on the fact that they are essentially at the same distance, concluded that the exponential convergence of the two spectra is not consistent with being due to the EBL photon-photon absorption only, and an intrinsic cutoff is required.
Our present analysis suggests that the pair-production effect predicted by our re-evaluated EBL intensity is sufficient to explain the VHE spectra for both sources, 
assuming intrinsic power-law shapes (that in the case of Mkn 421 are one unit, in the spectral index, steeper than for \object{Mkn 501}).


\section{Discussion and conclusions}

\subsection{Possible indications for new physics}

One of the important topics still open after several published analyses and long debate concerns the detailed spectral forms of the few local blazars whose Cherenkov spectra have been observed up to and above $E_0 \sim 10\ TeV$.
These analyses (see e.g., Horns \& Meyer 2012; Costamante 2013) reported evidence for substantial upturns in the VHE spectra of \object{Mkn 501} and \object{Mkn 421}, once corrected for pair-production absorption according to standard EBL models - like if such absorption corrections would be too large. Even our previous analysis in AF2008 did not exclude such an effect in the two examined local blazars. These inferred spectral upturns at several TeV to tens of TeV energies are not consistent with standard blazar emission models (except assuming rather ad-hoc effects of photon absorption and pile-up).
Indications for anomalies in the photon propagation and pair-production have been claimed by comparing the energy distributions for VHE photons received from distant cosmic sources (Horns \& Meyer 2012).

As discussed in several papers (e.g., De Angelis et al. 2007), this is an important issue for fundamental physics because, considering the very high energies of photons and particles involved, these effects might reveal deviations from the Standard Model and might require new physics.

The most obvious possibility for increasing the transparency of the universe would be to decrease the photon-photon collision cross-section as a consequence of a violation of the Lorentz invariance, LIV.
This may be a natural possibility in the framework of Quantum-Gravity theories, which predict a breakdown of classical physics at energies on the scale of the Planck energy (Jacob and Piran, 2008).
For example, a modification of the photon dispersion relation might lead to a shift of the energy threshold for pair production at the highest energies, and a decrease of the optical depth at increasing photon energy.

An alternative possibility considered for increasing the photon mean free path is to assume that photons oscillate into and from hypothetical axion-like particles (ALPs) when traveling in the presence of a magnetic field (e.g., De Angelis et al. 2007). During the ALP phase in their path to the Earth, photons would not suffer pair-production effects and would travel undisturbed, then increasing the universal transparency.

There may, however, no longer be a convincing need for new physics emerging from VHE observations of distant sources.
This request appears now to be alleviated by the present analysis because, compared with the results in AF2008, our re-analysis of the IR source emissivities at long IR wavelengths, with the corresponding slightly lower inferred EBL intensity values, imply somewhat reduced photon-photon opacities
at the highest photon energies at > 1 TeV, relaxing the claimed potential inconsistencies. 

The pair-production extinction corrections to the observed spectra of \object{Mkn 421} and \object{Mkn 501} reported in Fig. \ref{mkn} based on our best-guess EBL are consistent with simple spectral extrapolations from the lower photon energies.
The high-energy upturns in the corrected spectrum indicated in Costamante (2012) are not apparent here. Our current results are in line with those reported as fiducial models in Primack et al. (2011) and in Stecker et al. (2016). 
In our case, the modeling of the extragalactic source contribution to the EBL is purely empirical and based on deep photometric imaging data and the source luminosity functions at all wavelengths from UV to the sub-millimeter. Overall, our improved EBL modeling reduces, or even removes, the tension between VHE observations of blazars and standard physical interpretations of their spectra.

%
%


\begin{figure*}[!ht] 
\includegraphics[angle=0,width=0.99\textwidth,height=0.7\textwidth]{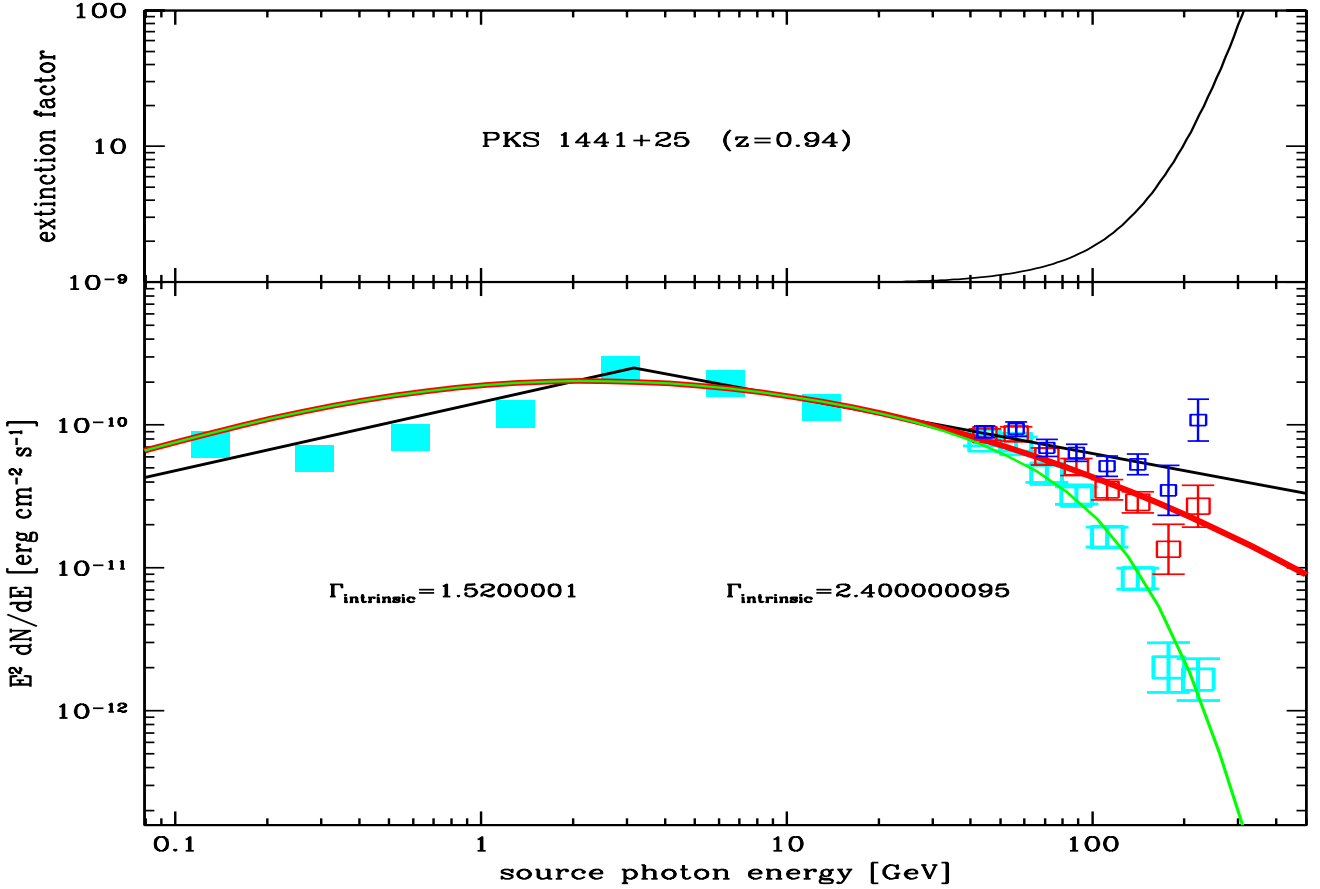}
\caption{
Spectral analysis of VHE observations of the {blazar} \object{PKS 1441+25}. Full cyan squares are from Fermi and open cyan squares are from MAGIC observations by Ahnen et al. (2015). The  data show a spectral steepening in the Fermi regime at $\epsilon\simeq 3$ GeV (the best-fit power laws are shown in black and the spectral indices in the insert) and an exponential cutoff in the range covered by MAGIC data. We fit these data with an intrinsic spectrum as for an LSP-ISP {blazar}, red curve, while the best-fitting green line includes an EBL absorption as from the present paper. 
Blue data points are the corrected spectrum assuming an optical/near-IR EBL intensity a factor 1.5 larger than our best-guess estimate.
The panel on top reports the absorption correction $exp[\tau_{\gamma\gamma}]$.
}
\label{1441}
\end{figure*}

\subsection{Revising the optical and near-IR EBL (COB) intensity}

Interactions between the highest-energy photons produced by cosmic sources and the background light will likely remain a hot topic in coming years.
While the pair production effect offers interesting constraints on the EBL intensity in the UV to the near-IR range, the exact level of this background is still subject to significant uncertainties. 

Zemcov et al. (2014), for example, report an observed excess signal of near-IR background fluctuations at 1.1 and 1.6 $\mu$m on scales of approximately 10 arc-minutes from a sounding rocket experiment, a signal that may correspond to diffuse light produced by stars in the intergalactic space outside galaxies, potentially implying an optical-NIR EBL around two times higher than our estimated galaxy contribution in Fig. 4. 

Abramowski et al. (2013) present a joint analysis of the spectra of a sample of {blazar}s totaling approximately $10^5$ VHE gamma-ray events. Further to the evidence of a clear, highly significant signature of the EBL absorption, the team reports an overall best-guess estimate for the EBL intensity over almost two decades of wavelengths from UV to near-IR with a peak value of $\nu I(\nu)\simeq 15\ nW/m^2/sr$ at 1.4 $\mu$, with essentially the same spectral shape of AF2008. This figure is slightly ($\sim$ 40\%) higher than our predicted value contributed by resolved sources as in Figure \ref{EBL}. The same analysis reveals that this excess flux corresponds to a pair-production optical depth larger than reported in AF2008 by a factor $\sim$1.34.

One of the interesting sources where one can look for EBL absorption effects is the high-redshift blazar \object{PKS1441+25} (z=0.94) that was detected with high significance by MAGIC during an outburst event in 2015 (Ahnen et al. 2015). We show in Figure \ref{1441} results of our analysis of the VHE spectrum of this source. The cyan data-points are the observed fluxes from Fermi and MAGIC, red datapoints are corrected for EBL absorption based on our best-guess optical depth. As can be seen in the Figure, the latter are well fit by a theoretical {blazar} spectrum (red curve, in $E^2 dN/dE$ units) with a peak energy at approximately 1 GeV,   a spectrum 
that is typical for the Intermediate or Low Synchrotron Peaked (ISP, LSP) objects. High-luminosity {blazar}s at large redshifts tend indeed to be of this category.

In the same figure, we also show (in blue) the MAGIC flux data corrected assuming an EBL intensity in the whole optical/near-IR a factor 1.5 larger than our best-guess estimate, following the suggestion by Abramowski et al., among others. These corrected data would be largely consistent with a power-law extrapolation (black line) of the lower-energy Fermi and MAGIC points, however implying a serious misfit of the highest-energy datum. How seriously this bad fit should be taken, and how physically consistent a power-law continuation of this kind, like the black line, could be (note that this is a high luminosity object, like ISP or LSP kind), cannot be decided with the present data.
Furthermore, it is difficult to be conclusive about the exact level of the EBL within the above mentioned boundaries.


\subsection{Future prospects}

Although atmospheric Cherenkov observatories, joining efforts with the Fermi all-sky surveys, are accumulating VHE photon detections at good rates from sources over a substantial range of redshifts, up to $z\sim 1$ as we have seen, it seems unlikely that this subject of photon propagation across the universe and EBL characterization will experience significant progress in coming years, further to already published results. This is due to the limited sensitivity, spectral resolution, and energy coverage of current instrumentation, on one side, and the complex interplay between physics and astrophysics on the other, as discussed in the previous subsection.

New-generation instrumentation is clearly needed to make the next step. One such occasion will be offered by the Cherenkov Telescope Array (CTA) in a few years from now (Actis et al. 2011). The full array will offer over factor 10 improvement in sensitivity and spatial and spectral resolution over current telescopes. Among the countless contributions to physics and astrophysics that can be envisaged from such an instrument, we certainly expect substantially improved constraints on the total EBL intensity (sources + diffuse emissions) over a wide interval of wavelengths from UV to the far-IR, as well as relevant constraints on its evolution with cosmic time.

As a particularly significant example of such an opportunity, we have considered using the pair-production absorption effect with an instrument like CTA to constrain the integrated emission of very high-redshift sources responsible for the cosmological re-ionization at redshifts of z$\sim$8-10 (that we name Population III, or Pop III sources henceforth). While partly dedicated to this purpose, the new generation space telescope JWST itself will not directly detect them if they are single stars or stellar aggregates.

Instead CTA might see their signatures in the spectra of high-redshift {blazar}s. This also takes advantage of the fact that, while the EBL contribution by known sources, such as galaxies, starts fading away at $z>1$ (see Figs. \ref{sfrEBL} and \ref{EBLnd}), the Pop III background proper photon number density increases very quickly and proportionally to $(1+z)^3$ up to z$\sim$7-10.

We have then performed simulations of future observations of a few tens of hours with CTA using the MAGIC and Fermi data of the z=0.94 {blazar} \object{PKS1441+25}  as a reference. We then assumed, on top of our modeled galaxy background, the existence of a truly diffuse signal, like that due to Pop III sources active at very high redshifts. For the latter, we adopted a spectrum as measured by Matsumoto et al. (2005), with a peak at 1.4 $\mu$m, but scaled down in flux, following a similar procedure as in AF2008 (see also Raue, Kneiske \& Mazin 2009).

\begin{figure*}[!ht] 
\includegraphics[angle=0,width=0.99\textwidth,height=0.7\textwidth]{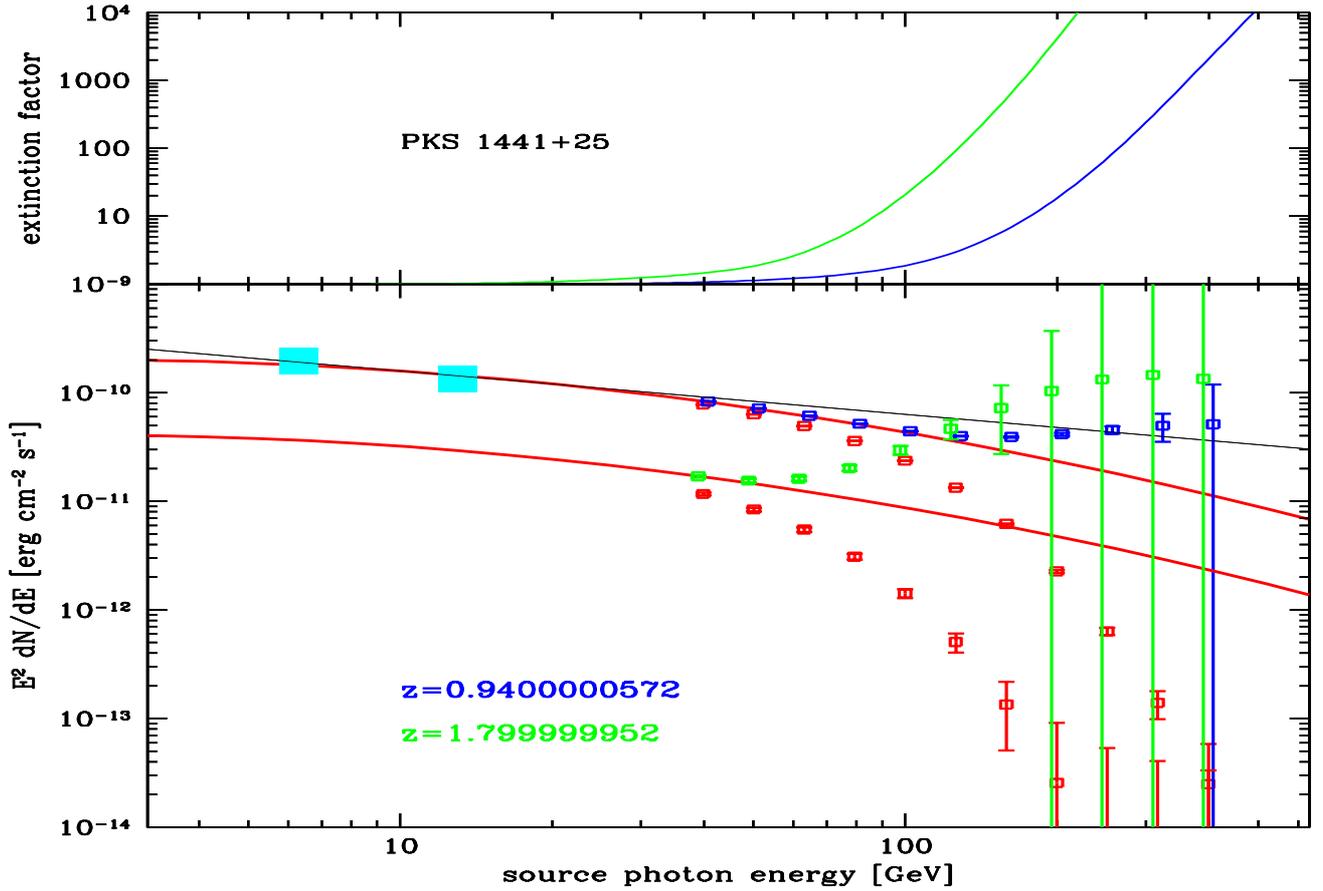}
\caption{
Similar to Fig. \ref{1441}, but including simulations of the performances of future observations with the CTA array (50 hrs observations) and different spectral corrections for photon-photon absorption. Cyan and red datapoints as in the previous figure. Here we assume the existence of a diffuse excess background in the near-IR from very high redshift sources (Population III objects), on top of our estimated EBL in the present paper. The figure includes a simulation of future observations of a similar gamma-ray source observed at the redshift z=1.8 with the CTA array, to show how sensitive the instrument will be in constraining truly diffuse background signals from un-resolved high-redshift source population. 
}
\label{1441simul}
\end{figure*}

We have simulated CTA observations of \object{PKS1441+25}, first of all assuming its measured redshift z=0.94, a luminosity as observed during the 2015 outburst, and a spectral shape as our best-fit red line in Fig. \ref{1441}. 
The simulation results are reported in Fig. \ref{1441simul}:
the upper continuous red line is the intrinsic source spectrum, while the upper sequence of red datapoints are the simulated data calculated assuming our best-guess EBL model for the photon-photon absorption.
The corresponding 1$\sigma$ error-bars are based on the predicted sensitivity and spectral resolution of the full CTA array for a 50-hour integration \footnote{CTA performances have been obtained from: \ https://portal.cta-observatory.org/Pages/Home.aspx}.

The blue data-points correspond to the simulated spectral data corrected for absorption assuming, in addition to the best-guess known-source contribution to the EBL, that of a Pop III component with an intensity equal to only 5\% of the IRTS background measured by Matsumoto et al. At the peak intensity wavelength of 1.4 $\mu$m, this corresponds to an intensity of 3.5 $nW/m^2/sr$, added to the known source contribution of 11.4 $nW/m^2/sr$ (see Fig. \ref{EBL}). This Pop III contribution would amount to 1.5 $nW/m^2/sr$ in bolometric units when integrated between 1 and 4 $\mu$m, corresponding to a comoving star-formation rate density of approximately 1 $M_\odot/yr/Mpc^3$ at $z\sim 10$, according to the analysis of Raue et al. (2009). As we see in Fig. \ref{1441simul}, this small truly diffuse component implies a moderate hardening of the (blue) corrected spectrum, however with an overall spectral shape not overly dissimilar to a simple power-law extrapolated from the lower-energy data shown as a black line.

The relative incidence of such a Pop III additional component would become much more evident for a 50-hour observation with CTA of the same source, assumed to be observed at z=1.8 with the same luminosity.  
To simulate this occurrence, we have scaled the spectrum of PKS1441 to that redshift and performed the same calculation as before.


The lower continuous red line in Figure \ref{1441simul} is the predicted intrinsic spectrum for z=1.8, while the lower red datapoints simulate the CTA observation including the EBL absorption for the known-source contribution only.
As we see, the source becomes fainter with z and, after application of the EBL absorption, it is detectable by CTA only up to 150 GeV, while the lower-redshift counterpart could be observed up to almost 400 GeV. 

We have then corrected back these simulated spectral data assuming the small additional contribution of Pop III sources as above (5\% of the IRTS background signal). The resulting absorption-corrected spectrum is shown as green datapoints in Figure \ref{1441simul}. We see that, in this case, the inclusion of a small Pop III signal dramatically increases the $\tau_{\gamma \gamma}$ for such a high-z source, as shown also by the comparison of the green and blue lines in the   upper panel.
At the photon energy of 200 GeV, the inclusion of the Pop III excess increases the $\tau_{\gamma \gamma}$ only by a factor 2 for a source at z=0.9 (see blue line in Fig. \ref{1441simul} versus the black one in Fig. \ref{1441}). Instead, for the same source observed at z=1.8, the inclusion of the small Pop III excess dramatically increases the optical depth, as shown by the green line compared to the blue one in the upper panel of Figure \ref{1441simul}. As mentioned, this is a consequence of the very fast $(1+z)^3$ increase of the Pop III photon density above z=1.
  This exercise illustrates how the low-z known source contributions to the EBL and residual high-z components could be disentangled  by comparing CTA observations of {blazar}s at different redshifts.

Our simulation exercise shows that new-generation Cherenkov arrays, with the vast expansion of the VHE cosmological horizon that they will offer, will provide us with new unique opportunities for sampling the history of radiant energy production in the universe.


\begin{acknowledgements}
We are glad to acknowledge useful discussions and exchanges with A. De Angelis, Mose' Mariotti, Michele Doro, Sara Buson, Abelardo Moralejo, and Daniel Mazin among others.
      This work was mostly supported by the University of Padova.
\end{acknowledgements}

\begin{table*}
      \caption[]{Photon proper number density as a function of redshift.}
         \label{tngamma}
\begin{tabular}{l l l l l l l l l l l l  }     
            \hline
            \noalign{\smallskip}
 \multicolumn{2}{c}{z=0}    & \multicolumn{2}{c} {z=0.2}   &
 \multicolumn{2}{c}{z=0.4}  & \multicolumn{2}{c} {z=0.6}   &   
 \multicolumn{2}{c}{z=0.8}  & \multicolumn{2}{c} {z=1.0}   \\
            \hline
$\log\epsilon$$^{\mathrm{a}}$&$\log(\epsilon\frac{dn_\gamma}{d\epsilon})$$^{\mathrm{b}}$   & $\log\epsilon$&$\log(\epsilon\frac{dn_\gamma}{d\epsilon})$   & $\log\epsilon$&$\log(\epsilon\frac{dn_\gamma}{d\epsilon})$   & $\log\epsilon$&$\log(\epsilon\frac{dn_\gamma}{d\epsilon})$   & $\log\epsilon$&$\log(\epsilon\frac{dn_\gamma}{d\epsilon})$   & $\log\epsilon$&$\log(\epsilon\frac{dn_\gamma}{d\epsilon})$   \\
            \noalign{\smallskip}
            \hline
            \noalign{\smallskip}
   -2.835  &  -0.9367 &  -2.756  &  -0.7423 & -2.689 &   -0.5311        &  -2.631 &   -0.3672 &  -2.58  &  -0.2245&  -2.534  &  -0.1062 \\
   -2.604  &  -0.3533 &  -2.525  &  -0.1642 & -2.458 &   0.04493        &    -2.4 &    0.2084 & -2.349  &   0.3406&  -2.303  &   0.4451 \\
    -2.45  &  -0.1126 &   -2.37  &  0.07846 & -2.303 &    0.2842        &  -2.245 &    0.4412 & -2.194  &   0.5546&  -2.148  &   0.6357 \\
   -2.303  & -0.03035 &  -2.224  &   0.1723 & -2.157 &    0.3701        &  -2.099 &    0.5147 & -2.048  &   0.6046&  -2.002  &   0.6636 \\
   -2.136  &  -0.0691 &  -2.057  &   0.1303 &  -1.99 &    0.3197        &  -1.932 &     0.449 & -1.881  &   0.5111&  -1.835  &   0.5391 \\
   -2.052  &  -0.1755 &  -1.972  &  0.01995 & -1.906 &    0.1956        &  -1.847 &    0.3064 & -1.796  &   0.3428&  -1.751  &   0.3508 \\
   -1.947  &  -0.3735 &  -1.868  &  -0.1889 & -1.801 &   -0.0317        &  -1.743 &   0.05994 & -1.692  &  0.08171&  -1.646  &  0.07664 \\
    -1.86  &  -0.5733 &   -1.78  &  -0.4037 & -1.714 &   -0.2571        &  -1.656 &   -0.1738 & -1.604  &  -0.1568&  -1.559  &  -0.1605 \\
    -1.78  &  -0.7724 &  -1.701  &  -0.6107 & -1.634 &   -0.4717        &  -1.576 &   -0.3885 & -1.525  &  -0.3697&   -1.48  &  -0.3711 \\
   -1.751  &  -0.8526 &  -1.671  &  -0.6934 & -1.604 &   -0.5541        &  -1.546 &   -0.4703 & -1.495  &  -0.4503&   -1.45  &  -0.4509 \\
   -1.684  &  -0.9834 &  -1.604  &  -0.8292 & -1.537 &   -0.6897        &   -1.48 &   -0.6048 & -1.428  &   -0.583&  -1.383  &   -0.583 \\
   -1.508  &   -1.418 &  -1.428  &   -1.265 & -1.361 &    -1.115        &  -1.303 &    -1.011 & -1.252  &  -0.9646&  -1.206  &   -0.942 \\
   -1.383  &   -1.634 &  -1.303  &   -1.469 & -1.236 &    -1.307        &  -1.178 &    -1.198 & -1.127  &   -1.146&  -1.082  &   -1.116 \\
   -1.303  &   -1.793 &  -1.224  &   -1.595 & -1.157 &    -1.431        &  -1.099 &    -1.318 & -1.048  &   -1.258&  -1.002  &   -1.242 \\
   -1.206  &   -1.886 &  -1.127  &   -1.691 &  -1.06 &    -1.521        &  -1.002 &    -1.412 &-0.9512  &   -1.351& -0.9055  &   -1.297 \\
   -1.082  &   -2.055 &  -1.002  &   -1.868 &-0.9355 &      -1.7        & -0.8775 &    -1.576 &-0.8262  &   -1.551& -0.7804  &   -1.589 \\
  -0.9846  &   -2.302 & -0.9055  &   -2.095 &-0.8386 &    -1.947        & -0.7804 &    -1.868 &-0.7293  &   -1.916& -0.6836  &   -1.993 \\
  -0.8598  &   -2.642 & -0.7804  &   -2.466 &-0.7135 &    -2.343        & -0.6556 &    -2.264 &-0.6045  &   -2.206& -0.5586  &   -2.171 \\
  -0.8085  &   -2.655 & -0.7293  &   -2.486 &-0.6623 &    -2.355        & -0.6045 &     -2.26 &-0.5533  &    -2.19& -0.5075  &   -2.144 \\
  -0.7314  &   -2.654 & -0.6523  &    -2.46 &-0.5854 &    -2.312        & -0.5274 &    -2.203 &-0.4763  &   -2.125& -0.4305  &   -2.079 \\
  -0.6688  &   -2.604 & -0.5897  &   -2.403 &-0.5227 &    -2.251        & -0.4647 &    -2.143 &-0.4136  &   -2.067& -0.3678  &   -2.018 \\
  -0.5586  &   -2.494 & -0.4795  &    -2.29 &-0.4125 &    -2.142        & -0.3546 &    -2.034 &-0.3034  &   -1.954& -0.2577  &   -1.913 \\
  -0.4618  &   -2.404 & -0.3826  &   -2.213 &-0.3156 &    -2.067        & -0.2577 &    -1.968 &-0.2065  &   -1.916& -0.1607  &   -1.903 \\
  -0.3826  &   -2.366 & -0.3034  &   -2.173 &-0.2364 &    -2.042        & -0.1785 &    -1.967 &-0.1273  &   -1.939&-0.08155  &   -1.948 \\
  -0.1359  &   -2.396 & -0.0567  &   -2.261 & 0.0103 &    -2.173        & 0.06819 &    -2.131 & 0.1193  &   -2.125&  0.1652  &   -2.148 \\
 -0.01936  &   -2.502 & 0.05994  &   -2.372 & 0.1268 &    -2.304        &  0.1847 &    -2.272 & 0.2358  &   -2.263&  0.2817  &   -2.283 \\
  0.09447  &   -2.644 &  0.1738  &   -2.544 & 0.2405 &    -2.478        &  0.2986 &    -2.451 & 0.3499  &   -2.462&  0.3955  &   -2.513 \\
   0.1915  &   -2.829 &  0.2707  &   -2.727 & 0.3377 &    -2.683        &  0.3955 &    -2.688 & 0.4467  &   -2.737&  0.4925  &   -2.811 \\
   0.3541  &   -3.147 &  0.4334  &   -3.114 & 0.5004 &    -3.106        &  0.5583 &    -3.065 & 0.6095  &   -3.021&  0.6552  &   -2.971 \\
   0.4925  &    -3.55 &  0.5717  &   -3.448 & 0.6386 &    -3.352        &  0.6966 &    -3.247 & 0.7477  &   -3.156&  0.7935  &   -3.085 \\
   0.7935  &   -4.175 &  0.8727  &   -4.038 & 0.9396 &    -3.971        &  0.9976 &    -3.954 &  1.049  &   -3.934&   1.094  &   -3.888 \\
            \noalign{\smallskip}
            \hline
         \end{tabular}
\begin{list}{}{}
\item[$^{\mathrm{a}}$] Photon energies $\epsilon$ are in eV.
\item[$^{\mathrm{b}}$] Photon densities $\epsilon\frac{dn_\gamma}{d\epsilon}$ are in $[cm^{-3}]$.
\end{list}
   \end{table*}
%

   \begin{table*}
      \caption[]{Photon proper number density as a function of redshift.}
         \label{tngamma1}
\begin{tabular}{l l l l l l l l l l }     
            \hline
            \noalign{\smallskip}
 \multicolumn{2}{c}{z=1.2}  & \multicolumn{2}{c} {z=1.4}   &
 \multicolumn{2}{c}{z=1.6}  & \multicolumn{2}{c} {z=1.8}   &
 \multicolumn{2}{c}{z=2.0}  \\
            \hline
$\log\epsilon$$^{\mathrm{a}}$&$\log(\epsilon\frac{dn_\gamma}{d\epsilon})$$^{\mathrm{b}}$   & $\log\epsilon$&$\log(\epsilon\frac{dn_\gamma}{d\epsilon})$   & $\log\epsilon$&$\log(\epsilon\frac{dn_\gamma}{d\epsilon})$   & $\log\epsilon$&$\log(\epsilon\frac{dn_\gamma}{d\epsilon})$   &
$\log\epsilon$&$\log(\epsilon\frac{dn_\gamma}{d\epsilon})$   \\
            \noalign{\smallskip}
            \hline
            \noalign{\smallskip}
   -2.492 & -0.009528   &   -2.455 &   0.06558  &   -2.42 &    0.1156   &  -2.388 &    0.1517     &   -2.358 &    0.1703 \\
   -2.262 &    0.5224   &   -2.224 &    0.5691  &  -2.189 &    0.5894   &  -2.157 &    0.5977     &   -2.127 &    0.5902 \\
   -2.107 &    0.6874   &   -2.069 &     0.703  &  -2.035 &     0.699   &  -2.002 &    0.6874     &   -1.972 &    0.6629 \\
   -1.961 &    0.6968   &   -1.923 &    0.6931  &  -1.888 &    0.6682   &  -1.856 &    0.6348     &   -1.826 &    0.5884 \\
   -1.793 &    0.5403   &   -1.756 &    0.5013  &  -1.721 &    0.4496   &  -1.689 &    0.3883     &   -1.659 &    0.3084 \\
   -1.709 &    0.3402   &   -1.671 &    0.2794  &  -1.637 &    0.2068   &  -1.604 &    0.1348     &   -1.574 &   0.05805 \\
   -1.604 &   0.05614   &   -1.567 & -0.003532  &  -1.532 &  -0.06778   &    -1.5 &   -0.1286     &    -1.47 &   -0.1946 \\
   -1.517 &    -0.177   &    -1.48 &   -0.2301  &  -1.445 &   -0.2892   &  -1.413 &   -0.3518     &   -1.383 &   -0.4279 \\
   -1.438 &   -0.3861   &     -1.4 &   -0.4394  &  -1.366 &   -0.4999   &  -1.333 &   -0.5591     &   -1.303 &   -0.6229 \\
   -1.408 &   -0.4646   &    -1.37 &   -0.5127  &  -1.336 &   -0.5672   &  -1.303 &   -0.6211     &   -1.273 &   -0.6834 \\
   -1.341 &   -0.5952   &   -1.303 &   -0.6364  &  -1.269 &   -0.6799   &  -1.236 &   -0.7261     &   -1.206 &   -0.7836 \\
   -1.165 &   -0.9404   &   -1.127 &   -0.9714  &  -1.093 &    -1.005   &   -1.06 &    -1.027     &    -1.03 &    -1.068 \\
    -1.04 &    -1.101   &   -1.002 &    -1.132  & -0.9678 &    -1.189   & -0.9355 &    -1.225     &  -0.9055 &    -1.228 \\
   -0.961 &    -1.237   &  -0.9234 &    -1.208  & -0.8884 &    -1.174   & -0.8564 &    -1.179     &  -0.8262 &     -1.24 \\
  -0.8642 &    -1.256   &  -0.8262 &    -1.274  & -0.7916 &    -1.343   & -0.7595 &    -1.451     &  -0.7293 &    -1.569 \\
  -0.7392 &     -1.69   &  -0.7014 &    -1.831  & -0.6666 &    -1.936   & -0.6343 &    -2.016     &  -0.6045 &    -2.058 \\
  -0.6423 &     -2.04   &  -0.6045 &    -2.088  & -0.5698 &    -2.081   & -0.5375 &    -2.068     &  -0.5075 &    -2.069 \\
  -0.5173 &    -2.158   &  -0.4795 &    -2.168  & -0.4448 &     -2.18   & -0.4125 &     -2.19     &  -0.3826 &    -2.215 \\
  -0.4661 &    -2.127   &  -0.4283 &    -2.135  & -0.3936 &    -2.143   & -0.3614 &    -2.144     &  -0.3314 &    -2.159 \\
  -0.3891 &    -2.057   &  -0.3513 &    -2.054  & -0.3166 &    -2.048   & -0.2844 &    -2.042     &  -0.2545 &    -2.066 \\
  -0.3265 &    -1.989   &  -0.2887 &    -1.979  & -0.2539 &    -1.985   & -0.2217 &    -2.001     &  -0.1918 &    -2.037 \\
  -0.2162 &    -1.911   &  -0.1785 &    -1.932  & -0.1437 &    -1.965   & -0.1115 &    -2.006     & -0.08155 &    -2.062 \\
  -0.1194 &    -1.928   & -0.08155 &    -1.967  &-0.04677 &    -2.005   &-0.01462 &    -2.041     &  0.01536 &     -2.09 \\
 -0.04015 &    -1.973   &-0.002395 &    -2.003  & 0.03222 &    -2.036   & 0.06446 &    -2.078     &  0.09447 &    -2.132 \\
   0.2066 &     -2.17   &   0.2443 &    -2.184  &   0.279 &    -2.204   &  0.3113 &    -2.234     &   0.3412 &    -2.272 \\
    0.323 &    -2.305   &   0.3608 &    -2.328  &  0.3955 &    -2.365   &  0.4278 &    -2.418     &   0.4577 &    -2.494 \\
    0.437 &    -2.569   &   0.4748 &    -2.632  &  0.5095 &    -2.683   &  0.5417 &    -2.706     &   0.5717 &    -2.724 \\
   0.5339 &    -2.827   &   0.5717 &    -2.818  &  0.6064 &     -2.81   &  0.6386 &    -2.803     &   0.6686 &    -2.801 \\
   0.6966 &    -2.907   &   0.7344 &    -2.846  &  0.7692 &    -2.798   &  0.8013 &    -2.764     &   0.8313 &    -2.749 \\
   0.8349 &    -3.028   &   0.8727 &    -2.989  &  0.9075 &    -2.969   &  0.9396 &    -2.978     &   0.9696 &    -3.022 \\
    1.136 &    -3.777   &    1.174 &    -3.666  &   1.208 &    -3.577   &   1.241 &    -3.531     &    1.271 &    -3.504 \\
            \noalign{\smallskip}
            \hline
         \end{tabular}
\begin{list}{}{}
\item[$^{\mathrm{a}}$] Photon energies $\epsilon$ are in eV.
\item[$^{\mathrm{b}}$] Photon densities $\epsilon\frac{dn_\gamma}{d\epsilon}$ are in $[cm^{-3}]$.
\end{list}
   \end{table*}

   \begin{table*}
      \caption[]{Photon-photon optical depth as a function of energy and redshift$^{\mathrm{a}}$.}
         \label{tautab}
\begin{tabular}{l l l l l l l l l l  }     
            \hline
            \hline
            \noalign{\smallskip}
        Energy     & $\tau(z,E_\gamma)$ & $\tau(z,E_\gamma)$ & $\tau(z,E_\gamma)$ & $\tau(z,E_\gamma)$ & $\tau(z,E_\gamma)$ & $\tau(z,E_\gamma)$ & $\tau(z,E_\gamma)$ & $\tau(z,E_\gamma)$ & $\tau(z,E_\gamma)$  \\
            \noalign{\smallskip}
    [TeV]      &   z=0.01                &      z=0.03         &  z=0.1                 &        z=0.3              &   z=0.5         &   z=1.0            &         z=1.5                &     z=2.0        &   z=3.0          \\
            \noalign{\smallskip}
            \hline
            \noalign{\smallskip}
 0.0200     & 2.91E-19           &   8.692E-19              & 2.758E-18         & 4.659E-07         & 1.167E-03   &   0.0545       &        0.2543        & 0.5830    &  1.206                        \\
 0.0240     & 2.91E-19               &   8.692E-19                  & 2.758E-18         & 8.606E-05             & 3.773E-03     &   0.0946       &        0.3564 & 0.7786          &  1.528                        \\
 0.0289     & 2.91E-19               &   8.692E-19                  & 2.758E-18         & 6.880E-04             & 9.205E-03     &   0.1390       &        0.4979 &  1.037          &  1.928                        \\
 0.0347     & 2.91E-19               &   8.692E-19                  & 4.388E-06         & 2.531E-03             & 1.899E-02     &   0.2100       &        0.6946 &  1.374          &  2.420                        \\
 0.0417     & 2.11E-07               &   2.832E-06                  & 1.523E-04         & 6.572E-03             & 3.506E-02     &   0.3138       &        0.9642 &  1.806          &  3.016                        \\
 0.0502     & 2.12E-05               &   9.445E-05                  & 8.587E-04         & 1.402E-02             & 6.016E-02     &   0.4659       &         1.325 &  2.352          &  3.730                        \\
 0.0603     & 1.13E-04               &   4.241E-04                  & 2.546E-03         & 2.635E-02             & 9.868E-02     &   0.6841       &         1.797 &  3.028          &  4.574                        \\
 0.0726     & 3.14E-04               &   1.103E-03                  & 5.605E-03         & 4.571E-02             & 0.1577        &   0.9877       &         2.400 &  3.847          &  5.556                        \\
 0.0873     & 6.64E-04               &   2.258E-03                  & 1.058E-02         & 7.572E-02             & 0.2476        &    1.396       &         3.151 &  4.817          &  6.681                        \\
 0.104      & 1.22E-03               &   4.097E-03                  & 1.849E-02         & 0.1221                & 0.3814        &    1.929       &         4.060 &  5.935          &  7.960                        \\
 0.126      & 2.12E-03               &   7.039E-03                  & 3.104E-02         & 0.1928                & 0.5738        &    2.602       &         5.129 &  7.194          &  9.406                        \\
 0.151      & 3.55E-03               &   1.167E-02                  & 5.045E-02         & 0.2971                & 0.8389        &    3.425       &         6.342 &  8.576          &  11.06                        \\
 0.182      & 5.74E-03               &   1.872E-02                  & 7.953E-02         & 0.4447                &  1.188        &    4.394       &         7.672 &  10.07          &  12.97                        \\
 0.219      & 8.96E-03               &   2.907E-02                  & 0.1216         & 0.6441                &  1.629        &    5.490       &         9.091 &  11.71          &  15.24                        \\
 0.263      & 1.36E-02               &   4.378E-02                  & 0.1800         & 0.9006                &  2.163        &    6.681       &         10.57 &  13.55          &  17.99                        \\
 0.316      & 1.98E-02               &   6.367E-02                  & 0.2573         &  1.214                &  2.781        &    7.928       &         12.13 &  15.70          &  21.35                        \\
 0.381      & 2.80E-02               &   8.935E-02                  & 0.3540         &  1.580                &  3.465        &    9.193       &         13.85 &  18.29          &  25.55                        \\
 0.458      & 3.79E-02               &   0.1205                 & 0.4683         &  1.988                &  4.187        &    10.46       &         15.83 &  21.47          &  31.00                        \\
 0.550      & 4.95E-02               &   0.1563                 & 0.5963         &  2.420                &  4.918        &    11.75       &         18.20 &  25.40          &  38.11                        \\
 0.662      & 6.22E-02               &   0.1953                 & 0.7337         &  2.858                &  5.630        &    13.14       &         21.10 &  30.39          &  47.92                        \\
 0.796      & 7.56E-02               &   0.2364                 & 0.8742         &  3.282                &  6.306        &    14.74       &         24.72 &  36.89          &  61.48                        \\
 0.957      & 8.89E-02               &   0.2768                 &  1.010         &  3.678                &  6.942        &    16.65       &         29.37 &  45.61          &  81.33                        \\
  1.15      & 0.101                      &   0.3152                     &  1.137          &  4.039                &  7.564        &    19.03       &          35.50        &  57.71          &  110.4                        \\
  1.38      & 0.113                      &   0.3501                     &  1.251          &  4.368                &  8.223        &    22.05       &          43.75        &  75.25          &  153.5                        \\
  1.66      & 0.123                      &   0.3810                     &  1.352          &  4.685                &  8.990        &    26.01       &          55.34        &  101.6          &  216.5                        \\
  2.000     & 0.132                      &   0.4076                     &  1.442          &  5.029                &  9.957        &    31.34       &          72.36        &  142.1          &  305.3                        \\
  2.40      & 0.140                      &   0.4318                     &  1.528          &  5.452                &  11.21        &    38.75       &          98.22        &  204.1          &  428.1                        \\
  2.89      & 0.147                      &   0.4560                     &  1.627          &  6.017                &  12.88        &    49.44       &          138.5        &  296.5          &  589.3                        \\
  3.47      & 0.157                      &   0.4863                     &  1.756          &  6.783                &  15.16        &    65.46       &          201.3        &  428.0          &  793.1                        \\
  4.17      & 0.170                      &   0.5284                     &  1.938          &  7.825                &  18.32        &    90.25       &          296.0        &  606.7          &  1040.                        \\
  5.02      & 0.188                      &   0.5887                     &  2.195          &  9.254                &  22.88        &    129.3       &          431.7        &  836.3          &  1328.                        \\
  6.03      & 0.215                      &   0.6732                     &  2.539          &  11.24                &  29.62        &    190.3       &          615.8        &  1115.          &  1649.                        \\
  7.26      & 0.250                      &   0.7847                     &  3.011          &  14.14                &  39.94        &    281.8       &          851.4        &  1439.          &  1998.                        \\
  8.73      & 0.301                      &   0.9447                     &  3.680          &  18.50                &  56.15        &    411.0       &          1134.        &  1796.          &  2361.                        \\
  10.5      & 0.373                      &    1.171                     &  4.662          &  25.23                &  81.67        &    582.1       &          1458.        &  2172.          &  2727.                        \\
  12.6      & 0.479                      &    1.513                     &  6.175          &  35.83                &  121.1        &    795.1       &          1806.        &  2546.          &  3080.                        \\
  15.2      & 0.646                      &    2.048                     &  8.516          &  52.53                &  179.6        &    1043.       &          2163.        &  2905.          &  3406.                        \\
  18.2      & 0.901                      &    2.871                     &  12.18          &  78.21                &  260.5        &    1315.       &          2505.        &  3235.          &  3699.                        \\
  21.9      &  1.30                      &    4.162                     &  17.95          &  115.9                &  365.0        &    1596.       &          2821.        &  3516.          &  3941.                        \\
  26.4      &  1.92                      &    6.177                     &  26.78          &  167.9                &  491.0        &    1869.       &          3090.        &  3742.          &  4129.                        \\
  31.7      &  2.86                      &    9.181                     &  39.64          &  234.1                &  633.0        &    2119.       &          3306.        &  3910.          &  4258.                        \\
  38.1      &  4.22                      &    13.47                     &  57.24          &  312.4                &  781.9        &    2330.       &          3462.        &  4016.          &  4325.                        \\
  45.8      &  6.05                      &    19.14                     &  79.29          &  398.3                &  927.3        &    2494.       &          3555.        &  4055.          &  4331.                        \\
  55.1      &  8.29                      &    26.00                     &  104.6          &  485.1                &  1059.        &    2604.       &          3585.        &  4034.          &  4278.                        \\
  66.2      &  10.8                      &    33.59                     &  131.5          &  566.2                &  1168.        &    2658.       &          3555.        &  3958.          &  4168.                        \\
  79.6      &  13.4                      &    41.42                     &  158.0          &  635.6                &  1248.        &    2659.       &          3470.        &  3829.          &  4013.                        \\
  95.7      &  15.9                      &    48.81                     &  181.3          &  688.3                &  1294.        &    2609.       &          3340.        &  3657.          &  3816.                        \\
  115.      &  18.1                      &    54.98                     &  199.9          &  721.9                &  1307.        &    2518.       &          3171.        &  3451.          &  3587.                        \\
  138.      &  19.9                      &    59.82                     &  213.4          &  735.3                &  1290.        &    2393.       &          2974.        &  3220.          &  3338.                        \\
  166.      &  21.0                      &    63.03                     &  220.5          &  729.6                &  1247.        &    2244.       &          2759.        &  2975.          &  3076.                        \\
            \hline
            \noalign{\smallskip}
            \hline
         \end{tabular}
\begin{list}{}{}
\item[$^{\mathrm{a}}$] The absorption multiplicative factor is $e^{-\tau(z,E_\gamma)}$.
\end{list}
   \end{table*}

\end{document}